\documentclass[12pt]{iopart}
\usepackage{iopams}
\usepackage{amstext}
\usepackage{amsthm}
\usepackage{dsfont}
\usepackage{hyperref} 
\usepackage{enumitem}
\expandafter\let\csname equation*\endcsname=\relax
\expandafter\let\csname endequation*\endcsname=\relax
\usepackage{amsmath}
\usepackage{caption}
\usepackage{tikz}
\usepackage{subcaption}
 
\usepackage{graphicx}

\newcommand{\diff}{{\rm d}}
\newcommand{\EE}{\mathbb{E}}
\newcommand{\TT}{\mathbb{T}}
\renewcommand{\BB}{\mathbb{B}}
\renewcommand{\AA}{\mathbb{A}}
\newcommand{\tL}{\tilde{L}}
\newcommand{\tE}{\tilde{E}}
\newcommand{\tF}{\tilde{F}}
\newcommand{\torus}{\mathbb{T}_L^d}
\newcommand{\doubletorus}{\TT_{2L}^d}

\newcommand{\boxL}{\mathbb{B}_L^d}
\newcommand{\doubleboxL}{\mathbb{B}_{2L}^d}

\newcommand{\var}{\mathrm{var}}
\newcommand{\PP}{\mathbb{P}}
\newcommand{\NN}{\mathbb{N}}
\newcommand{\ZZ}{\mathbb{Z}}
\newcommand{\RR}{\mathbb{R}}
\newcommand{\posint}{\ZZ_{+}}
\newcommand{\naturals}{\NN}
\newcommand{\sA}{\mathcal{A}}
\newcommand{\sB}{\mathcal{B}}
\newcommand{\sC}{\mathcal{C}}
\newcommand{\sE}{\mathcal{E}}
\newcommand{\sK}{\mathcal{K}}

\newcommand{\sN}{\mathcal{N}}

\newcommand{\sS}{\mathcal{S}}
\newcommand{\sT}{\mathcal{T}}

\newcommand{\sX}{\mathcal{X}}
\newcommand{\Xgen}{\sX^{\ast}}
\newcommand{\XP}{\sX^{\mathrm{P}}}
\newcommand{\XR}{\sX^{\mathrm{R}}}
\newcommand{\XH}{\sX^{\mathrm{H}}}
\newcommand{\sZ}{\mathcal{Z}}
\newcommand{\rP}{\mathrm{P}}
\newcommand{\rR}{\mathrm{R}}
\newcommand{\rH}{\mathrm{H}}
\newcommand{\dc}{d_{\rm c}}
\newcommand{\ind}{{\mathds{1}}}
\newcommand{\erfc}{\mathrm{erfc}}

\newcommand{\meanNL}{\mathbb{E}(\mathcal{N}_L)}

\newcommand{\pnbar}{\bar{p}_n}

\newcommand{\zc}{z_{\mathrm{c}}}
\newcommand{\abs}[1]{\left\lvert #1 \right\rvert}
\newcommand{\floor}[1]{\lfloor #1 \rfloor}
\newcommand{\ceil}[1]{\lceil #1 \rceil}

\newtheorem{theorem}{Theorem}[section]
\newtheorem{lemma}[theorem]{Lemma}

\newtheorem{proposition}[theorem]{Proposition}

\newcommand{\dt}{\diff t}

\begin{document}

\title{Two-point functions of random-length random walk on high-dimensional boxes}

\author{Youjin Deng$^{1,2}$, Timothy M. Garoni$^3$, Jens Grimm$^3$, Zongzheng Zhou$^3$}
\address{$^1$Department of Modern Physics, University of Science and
  Technology of China, Hefei 230026, China}
\address{$^2$Hefei National Laboratory,
University of Science and Technology of China, Hefei, Anhui 230088, P.R. China}
\address{$^3$ School of Mathematics, Monash University, Clayton, Victoria 3800, Australia}
\ead{\mailto{tim.garoni@monash.edu}, \mailto{eric.zhou@monash.edu}, \mailto{yjdeng@ustc.edu.cn}}

\maketitle
 
\begin{abstract}
We study the two-point functions of a general class of random-length random walks on finite boxes in $\ZZ^d$ with $d\ge3$, and provide precise
asymptotics for their behaviour. We show that the finite-box two-point function is asymptotic to the
infinite-lattice two-point function when the typical walk length is $o(L^2)$, but develops a plateau when the typical walk length is
$\Omega(L^2)$. We also numerically study walk length moments and limiting distributions of the self-avoiding walk and Ising model on
five-dimensional tori, and find that they agree asymptotically with the known results for self-avoiding walk on the complete graph, both at
the critical point and also for a broad class of scaling windows/pseudocritical points. Furthermore, we show that the two-point function of
the finite-box random-length random walk, with walk length chosen via the complete graph self-avoiding walk, agrees numerically with the
two-point functions of the self-avoiding walk and Ising model on five-dimensional tori. We conjecture that these observations in five
dimensions should also hold in all higher dimensions.
\end{abstract}


\section{Introduction}
\label{sec:introduction}
The effects of boundary conditions on the finite-size scaling of statistical-mechanical lattice models in high dimensions has
a rather long history~\cite{BrezinZinnJustin1985,Binder1985,BinderNauenbergPrivmanYoung1985,LuijtenBloete1997,LuijtenBinderBloete99},
but has remained a very active area; see
e.g.~\cite{Lundow2021,LundowMarkstrom2011,FloresSolaBercheKennaWeigel2016,Lundow2015complete,LundowMarkstrom2016,GrimmElciZhouGaroniDeng2017,WittmannYoung2014,BercheKennaWalter2012,LundowMarkstrom2014}. One 
particular topic of interest has been the scaling of the Ising susceptibility at the infinite volume critical point, where it has been
observed numerically that on boxes of side-length $L$, periodic boundary conditions produce a scaling $L^{d/2}$, in contrast to the $L^2$
behaviour observed with free boundary conditions.  

Significant progress explaining this phenomenon mathematically was recently presented in~\cite{MichtaParkSlade2023}, where a rigorous
renormalisation group analysis was performed of the weakly coupled hierarchical $|\varphi|^4$ model in $d\ge4$ dimensions, on finite boxes
of volume $L^d$ with both periodic and free boundary conditions. In particular, it was shown that while the effective critical point for the 
model with periodic boundary conditions coincides with the infinite volume critical point, the effective critical point for free boundary
conditions is shifted away from the infinite-volume value by an amount of order $L^{-2}$. Moreover, it was shown that for both boundary
conditions, increasing the temperature above the effective critical point by an amount of order $L^{-\lambda}$ leads the susceptibility to
scale as $L^\lambda$, for any $3/2 \le \lambda < d/2$, but as $L^{d/2}$ for any $\lambda\ge d/2$. By universality, one would expect the same
behaviour to hold for the self-avoiding walk (SAW) and Ising model on boxes in $\ZZ^d$, and indeed numerical evidence
supporting this belief has been presented in~\cite{ZhouGrimmFangDengGaroni2018}. 

Another striking, and related, feature of high-dimensional models with periodic boundary conditions is the so-called plateau which emerges in
their two-point function, sufficiently close to the critical point, so that the initial simple random walk decay $|x|^{2-d}$ becomes
subdominant to a term which is independent of $x$ but decaying in $L$. For the Ising model on tori $\torus$ with $d>4$, it was proved
in~\cite{Papathanakos2006} that at the (infinite volume) critical point the two-point function is bounded below by
$c_1|x|^{2-d}+ c_2 L^{-d/2}$, and it was conjectured that an upper bound of the same order should exist.
This conjecture was extended in~\cite{ZhouGrimmFangDengGaroni2018}, where it was predicted that for the 
SAW and Ising model on $\torus$ with $d>4$, at temperatures shifted above the infinite volume critical point by $L^{-\lambda}$, the
two-point function behaves as $c_1|x|^{2-d}+ c_2 L^{\lambda-d}$ when $2\le \lambda \le d/2$, and as $c_1|x|^{2-d}+ c_2 L^{-d/2}$ when
$\lambda \ge d/2$. The latter behaviour has recently been established rigorously~\cite{Slade2023} for the Domb-Joyce model with
$d>4$, for sufficiently weak interaction strength. Moreover, analogous behaviour is also now known for bond
percolation~\cite{HutchcroftMichtaSlade2023} when $d\ge11$ for the nearest-neighbour model, and $d>6$ for spread-out models. We refer to the
regime with $0< \lambda < d/2$ as the high-temperature scaling window, and to the regime $\lambda> d/2$ as the critical window. It was
observed numerically in~\cite{ZhouGrimmFangDengGaroni2018} that for the SAW and Ising model with $\lambda<2$, the two-point function decays
faster than a power-law on large scales, but no conjecture for the precise nature of this behaviour was made.  

The general plateau behaviour conjectured in~\cite{ZhouGrimmFangDengGaroni2018} was supported by numerical simulations of the SAW and Ising
model on five-dimensional tori, but was motivated by considering a model of simple random walk in which the walk length is chosen to be
finite and random, and distributed as a SAW on the complete graph. The behaviour of SAW on the complete graph has been recently
studied~\cite{Yadin2016,DengGaroniGrimmNasrawiZhou2019,Slade2020}. Yet the behaviour of the two-point function of random-length random walk,
with arbitrary walk length distributions, appears not to have been studied in significant detail. This question was recently addressed for
random-length random walk on $\ZZ^d$ in~\cite{DengGaroniGrimmZhou2022}. One contribution of the current work is to present sharp asymptotic
results for the two-point function of random-length random walks on finite boxes in $\ZZ^d$, for three distinct choices of boundary
conditions, and with only modest assumptions on the walk length distribution. In summary, we find that if the walk length is concentrated on
a scale $o(L^2)$, then the finite-box and infinite lattice two-point functions are asymptotic. This is to be expected, since simple random
walks of length $N$ typically explore distances of order $\sqrt{N}$. By contrast, for walks whose expected length is $\Omega(L^2)$, we
establish a plateau given by the ratio of the mean walk length to the system volume. Specialising to the case that the walk length
distribution is that of SAW on the complete graph, we find a universal model of high-dimensional torus two-point function behaviour, that
agrees numerically with the SAW and Ising model on five-dimensional tori, in both the critical window and the high-temperature scaling
window for any $0<\lambda<d/2$. Moreover, we find that the high-temperature scaling window consists of two separate regimes; universal
exponential decay in terms of the continuum Green function for $\lambda<2$, and $\lambda$-dependent plateau for $2<\lambda<d/2$.

We note that in the special case in which the walk length is geometrically distributed, the two-point function of random-length random walk
on $\ZZ^d$ corresponds to the lattice Green function, which is very well studied; see~\cite{MichtaSlade2022} and references therein. In
that case, random-length random walk is generally referred to as \emph{killed} random walk~\cite{LawlerLimic2010}. We also note that plateau
behaviour of the geometrically-killed simple random walk two-point function on tori was recently established in~\cite[Theorem
  1.4]{Slade2023}, as a corollary of their weakly self-avoiding walk result. The results we present here for random-length random walk are both
sharper and more general than given in~\cite[Theorem 1.4]{Slade2023}.

The key assumption underlying the use of the complete graph SAW length in the random-length random walk to describe the SAW and Ising
two-point functions on $\torus$, is that for $d>4$, the large $L$ behaviour of the length of the SAW or Ising walk on $\torus$ should behave
in the same way as SAW on the complete graph; see Section~\ref{sec:saw_ising} for a definition of the Ising walk. We therefore now summarise
the known behaviour~\cite{Yadin2016,DengGaroniGrimmNasrawiZhou2019,Slade2020} of the SAW on the complete graph, $K_n$. At fugacities
$1/n(1+a n^{-p})$, the mean walk length scales as $n^{-p}$ for $p\in(0,1/2)$ and $a>0$, but scales as $\sqrt{n}$ for all $p\ge 1/2$ and
$a\in\RR$. Analogous behaviour has also been established~\cite{Slade2021} for SAW on the hypercube, $\ZZ_2^N$, and for weakly self-avoiding
walk on the torus $\torus$ with $d>\dc$ when the interaction strength is sufficiently small~\cite{MichtaSlade2023}. Moreover, the variance
and limiting distributions of the appropriately scaled/standardised length of SAW on $K_n$ are also known in
detail~\cite{DengGaroniGrimmNasrawiZhou2019,Slade2020}. For $p\in(0,1/2)$ with $a>0$ the variance scales as $n^{2p}$, and the walk length
divided by its mean converges to a mean-1 exponential distribution, while for $p>1/2$ the variance scales as $n$ and the standardised walk
length converges to a half-normal distribution. In Section~\ref{subsec:universal walk length distributions} we provide strong numerical
evidence that the same behaviours hold for the SAW and Ising walk length on $\torus$ with $d=5$, and we conjecture that they in fact hold
for all $d\ge5$.

\subsection{Outline}
Let us outline the remainder of this article. Section~\ref{subsec:notation} lists some notational conventions. 
Section~\ref{sec:saw_ising} defines the specific quantities of interest for the SAW and Ising model. 
Section~\ref{sec:random walk models} describes the random-length random walk models considered, and presents our main results for their
two-point functions. Section~\ref{sec:numerical results} then presents numerical results for the SAW and Ising model on five-dimensional
tori, both in the critical window and high-temperature scaling window. Specifically, Section~\ref{subsec:universal walk length
  distributions} presents numerical results for the SAW and Ising walk length, while Section~\ref{subsec:universal two-point functions}
considers their two-point functions. Finally, Sections~\ref{sec:proof fast decay} and~\ref{sec:proof plateau} present proofs of
Propositions~\ref{theorem:two-point function fast decay} and \ref{theorem:two-point function plateau}, respectively, and
Section~\ref{sec:proof of projection lemma} presents a proof of Lemma~\ref{lemma:two-point function projection identities}.

\subsection{Notation}
\label{subsec:notation}
For integer $d\ge1$ and $L>2$, we let $\boxL:=[-L/2,L/2)^d\cap\ZZ^d$. For each $x \in \boxL$, we denote its Euclidean norm by
$|x|:=\sqrt{x\cdot x }$. We let $\torus$ denote the $d$-dimensional discrete torus, of linear size $L$. We view $\torus$ both as a graph,
whose vertex set is taken to be $\boxL$, and also, when convenient, as a module over the commutative ring $\TT_L:=[-L/2,L/2)\cap\ZZ$
in which addition and multiplication are defined modulo $L$.

The standard asymptotic symbols such at $O$, $o$ etc will refer to large $L$ asymptotics. The definition of random-length random walk
requires a sequence of random walk lengths, $(\sN_L)_L$. In the asymptotic results we present for random-length random walk, the implied
constants may depend on $d$ and the choice of the sequence of distributions corresponding to $(\sN_L)_L$. Statements such as $f=O(g)$ in
that context then mean that for any particular choice of $d$ and the sequence of walk length distributions, there exists a constant $c>0$
such that $f(L)\le c\, g(L)$ for all sufficiently large $L$. In the case that constants depend on additional parameters, we will highlight
this via subscripts; e.g. if for fixed $\lambda$ we have $f(L,\lambda)\le c(d,\lambda) g(L,\lambda)$ for all $L\ge N(d,\lambda)$, then we
will write $f=O_\lambda(g)$. If $f=O(g)$ and $g=O(f)$ we write $f\asymp g$. We find it convenient to also use the Vinogradov symbols, so
that $f\ll g$ is equivalent to $f=O(g)$, and $f\gg g$ is equivalent to $g=O(f)$. We also find it convenient to write $f=\Omega(g)$ to denote
$g=O(f)$ and $f=\omega(g)$ to denote $g=o(f)$. 

The set of non-negative integers will be denoted by $\naturals$, and $\posint:=\naturals\setminus0$. For any $n\in\posint$ we write
$[n]:=\{1,2,\ldots,n\}$. 

\section{The SAW and Ising model}
\label{sec:saw_ising}
Let $G=(V,E)$ be a rooted graph, with root $0$. For $n\in\naturals$, let $\Omega_G^n$ denote the set of all $n$-step walks on $G$ which
start at $0$; i.e. all sequences $\omega_0,\ldots,\omega_n$ such that $\omega_i\in V$, $\omega_0=0$ and $\omega_i\,\omega_{i+1}\in E$. We
set $\Omega_G:=\bigcup_{n\in\naturals}\Omega_{G}^n$. For $\omega\in \Omega_{G}^n$, the notation $\omega:0\to v$ implies $\omega_n=v$, and we
denote the end of $\omega$ by $e(\omega)=\omega_n$. In what follows, we let $|\omega|$ denote the number of steps, or \emph{length}, of the
walk $\omega\in\Omega_G$, so that 
\begin{equation}
  |\omega|=n \text{ iff $\omega\in \Omega_{G}^n$}. \label{eq:walk length definition}
\end{equation}
A walk $\omega\in\Omega_G$ is \emph{self-avoiding} if $\omega_i\neq\omega_j$ for all $i\neq j$. We consider the variable-length ensemble of
self-avoiding walks on $G$, and let $\sS$ denote a random SAW with this distribution, so that for all $\omega\in\Omega_G$
\begin{equation}
  \PP(\sS=\omega) = \frac{\rho(\omega)}{\sum_{\omega'\in\Omega_G}\rho(\omega)}\;,
\end{equation}
with 
\begin{equation}
    \rho(\omega) = z^{|\omega|}\ind(\text{$\omega$ is self-avoiding})\;.
  \label{eq:SAW weight}
\end{equation}
The quantity $z>0$ is the \emph{fugacity}. We will be interested in the distribution of the walk length $|\sS|$, and the 
two-point function defined by~\cite{MadrasSlade1996}
\begin{align}
    g(x)&:=\sum_{\substack{\omega\in \Omega_G \\ \omega:0\to x}} \rho(\omega)\;,   \label{eq:SAW two-point function} \\
    &= \frac{\PP[e(\sS)=x]}{\PP[|\sS|=0]}\;. \label{eq:S estimator of SAW g}
\end{align}
Our simulations of $\sS$, discussed below, were performed using a lifted version~\cite{HuChenDeng2017} of the Berretti-Sokal
algorithm~\cite{BerrettiSokal1985}.

We also consider analogous quantities for the Ising model. The zero-field ferromagnetic Ising model on finite graph $G=(V,E)$ at inverse
temperature $\beta\ge0$ is defined by the measure
\begin{equation}
  \PP(\sigma) \propto \exp\left(\beta \sum_{ij\in E} \sigma_i\sigma_j\right), \qquad \sigma\in \{-1,1\}^V.
  \label{Ising measure}
\end{equation}
The corresponding two-point function is defined by
\begin{equation}
  g(x):= \EE (\sigma_0\,\sigma_x)\;.
  \label{eq:Ising g definition}
\end{equation}
The Ising two-point function can be conveniently re-expressed via the high-temperature expansion, as follows. For $v\in V\setminus 0$, let
$\sC_v$ denote the set of all $A\subseteq E$ such that the set of all vertices of odd degree in $(V,A)$ is precisely $\{0,v\}$, and let
$\sC_0$ denote the set of all $A\subseteq E$ such that $(V,A)$ has no vertices of odd degree. For a family of edge sets $S\subseteq 2^E$,
let
\begin{equation}
  \lambda(S):=\sum_{A\in S} [\tanh(\beta)]^{|A|}.
\end{equation}
By analogy with the SAW case, we refer to $z=\tanh(\beta)$ as the Ising fugacity.
The high-temperature expansion for the Ising model (see e.g. \cite[(3.5)]{Aizenman1985} or~\cite[Lemma
  2.1]{CollevecchioGaroniHyndmanTokarev2016}) implies that we can re-express~\eqref{eq:Ising g definition} so that for all $x\in V$
\begin{equation}
  g(x) = \frac{\lambda(\sC_{x})}{\lambda(\sC_0)}\;.
  \label{eq:Ising high temperature}
\end{equation}
This high-temperature representation of the Ising model can also be used to provide a natural definition of the \emph{Ising walk},
first discussed in~\cite{Aizenman1982,Aizenman1985}, and studied numerically in~\cite{DengGaroniGrimmZhou2022}. 
Fix an (arbitrary) ordering, $\prec$, of $V$. We define $\mathcal{T}:\cup_{v\in V}\sC_{v}\to\Omega_G$ as follows.
If $A\in \sC_0$, then $\sT(A)=0$. If $A \in \sC_{v}$ with $v\neq 0$, we recursively define
the walk $\sT(A) = v_0v_1...v_k$ from $v_0 = 0$ to $v_k = v$, such that from $v_i$ we choose $v_{i+1}$ to be the
smallest neighbour of $v_i$ such that $v_iv_{i+1} \in A$ and $v_i v_{i+1}$ has not previously been traversed by the walk. 
It is clear that $\mathcal{T}(A)$ defines an edge self-avoiding trail from $0$ to $v$.

Now let $\sA$ denote a random element of $\cup_{v\in V}\,\sC_{v}$ with distribution
\begin{equation}
  \PP(\sA=A) = \frac{z^{|A|}}{\sum_{A'\in\cup_{v\in V}\,\sC_v}\,z^{|A'|}}\;,\qquad A\in\cup_{v\in V}\,\sC_{v}\;.
\end{equation}
The distribution of $\sA$ is precisely the stationary distribution of the Prokofiev-Svistunov worm algorithm~\cite{ProkofevSvistunov2001},
in which the worm tail is fixed at the root. Our simulations of $\sA$, discussed below, were performed using such a worm algorithm. We will
be interested in the induced distribution of $\sT:=\sT(\sA)$, and particularly in the distribution of its length, $|\sT|$, which we refer to
as the \emph{Ising walk length}.

We note that by partitioning $\Omega_G$ in terms of $\sT$, we can re-express the Ising two-point function~\eqref{eq:Ising g definition} in
precisely the form~\eqref{eq:SAW two-point function} but with  
\begin{equation}
  \rho(\omega)= \frac{\lambda(\sT^{-1}(\omega))}{\lambda(\sC_0)}.
  \label{eq:Ising two-point weight}
\end{equation}
Moreover, it can also be re-expressed in the form~\eqref{eq:S estimator of SAW g} with $\sS$ replaced by $\sT$.

The simulations of the SAW and Ising model to be presented in Section~\ref{sec:numerical results} were performed on five-dimensional
tori. As we will demonstrate, the asymptotic behaviour of both $|\sS|$ and $|\sT|$ appear to coincide with the
known~\cite{DengGaroniGrimmNasrawiZhou2019,Slade2020} asymptotic behaviour of $|\sS|$ on the complete graph, which we now summarise. Let $G$
be the complete graph $K_n$, rooted at a fixed vertex, and suppose the fugacity $z$ satisfies $1/z = n(1+a n^{-p})$. Let $\sK$ denote $\sS$
in this setting. It is known~\cite{Yadin2016} that the critical fugacity is $z=1/n$. Moreover, if $p<1/2$ and $a>0$ then we have for large
$n$ that 
  \begin{equation}
  \EE(|\sK|) \sim \frac{n^{p}}{a}\;, \qquad\qquad
  \var(|\sK|) \sim\left(\frac{n^{p}}{a}\right)^2 
  \label{eq:Kn SAW cumulants - high temperature scaling window}
  \end{equation}
  while if $p>1/2$ 
  \begin{equation}
  \EE(|\sK|) \sim \sqrt{\frac{2}{\pi}} \sqrt{n}\;, \qquad\qquad
  \var(|\sK|) \sim \left(1-\frac{2}{\pi}\right) n.
  \label{eq:Kn SAW cumulants - critical window}
\end{equation}
Furthermore, let $X$ be a standard normal random variable, and let $Y$ be an exponential random variable with mean 1. Then as $n\to\infty$ we
have for $p<1/2$ and $a>0$ that
  \begin{equation}
    \frac{|\sK|}{\EE(|\sK|)} \implies Y\;,
    \label{eq:scaling window Kn limit theorem}
  \end{equation}
while if $p>1/2$ then
  \begin{equation}
    \frac{|\sK|-\EE(|\sK|)}{\sqrt{\var(|\sK|)}} \implies \frac{|X|-\EE(|X|)}{\sqrt{\var(|X|)}}\;.
    \label{eq:critical window Kn limit theorem}
  \end{equation}
For later reference, we shall denote by $F$ the law of the standardised version of $|X|$, i.e. for $x\in\RR$
\begin{equation}
  F(x):= \PP\left(\frac{|X|-\EE(|X|)}{\sqrt{\var(|X|)}}\le x\right).
  \label{eq:def of standardised half normal}
\end{equation}
  
\subsection{Numerical details}
Our simulations of the SAW and Ising model were performed on 5-dimensional tori, at pseudocritical points $z_L = \zc(1- L^{-\lambda})$ for
various $\lambda >0$, where $\zc$ denotes the estimated location of the infinite-volume critical point. In the Ising case we used the
estimate $\zc = 0.113~424~8(5)$~\cite{LundowMarkstrom2014}, while in the SAW case we used
$z_{\mathrm{c}}=0.113~140~84(1)$~\cite{HuChenDeng2017}. A detailed analysis of integrated autocorrelation time is presented
in~\cite{DengGaroniSokal2007} for the worm algorithm and in~\cite{HuChenDeng2017} for the lifted Berretti-Sokal algorithm. Our fitting
methodology and corresponding error estimation follow standard procedures, see for instance~\cite{Young2015,Sokal1997}. 

\section{Random walk models}
\label{sec:random walk models}
\subsection{Definitions and boundary conditions}
Let $(C_n)_{n \in \naturals}$ be an i.i.d.~sequence of uniformly random elements of $\{\pm e^1,...,\pm e^d\}$, where
$e^i=(0,\ldots,1,\ldots,0)\in\ZZ^d$ is the standard unit vector along the $i$th coordinate axis, and 
let $\sN$ be an $\naturals$-valued random variable independent of $(C_n)_{n \in \naturals}$.
The corresponding random-length random walk (RLRW) on $\ZZ^d$ is the process $\sZ:=(\sZ_t)_{t=0}^\sN$ defined so that $\sZ_0=0$ and
$\sZ_{t}=\sZ_{t-1}+ C_{t}$ for each $1\le t \le \sN$. We also consider RLRWs $(\XP_t)_{t=0}^{\sN}$, $(\XR_t)_{t=0}^{\sN}$, and
$(\XH_t)_{t=0}^{\sN}$ on $\boxL$, with periodic, reflecting, and holding boundary conditions, respectively, defined so that $\Xgen_0=0$, and
for all $1\le t \le \sN$ we have $\Xgen_{t}=\Xgen_{t-1}+C_{t}$ if $\Xgen_{t-1}+C_{t}\in\boxL$, otherwise
\begin{align}
\label{Eq:Random walk on torus}
\XP_{t}&:= \XP_{t-1}+C_{t}(1-L)
\\
\label{eq:reflecting random walk}
\XR_{t}&:= \XR_{t-1}-C_{t}
\\
\label{eq:holding random walk}    
\XH_{t}&:= \XH_{t-1}
\end{align}
when $\Xgen_{t-1}+C_{t}\not\in\boxL$, where $\ast$ denotes either P, R or H, as appropriate.

We define the two-point function of $\sZ$ to be
\begin{equation}
  g_{\sN}(x):=\mathbb{E}\Bigg(\sum_{t=0}^{\sN} \ind(\sZ_t= x)\Bigg), \qquad x\in\ZZ^d\;.
\label{def:greens_function_rlrw}
\end{equation}
As noted in the Introduction, in the special case in which $\sN$ is geometrically distributed, the two-point function of RLRW on $\ZZ^d$
corresponds to the lattice Green function. Analogous definitions hold for $\XP$, $\XR$, $\XH$. Specifically, for such a process on $\boxL$ we set
\begin{equation}
    g_{\ast,\,L,\,\sN}(x):=\mathbb{E}\Bigg(\sum_{t=0}^{\sN} \ind(\Xgen_t= x)\Bigg)\;,\qquad x\in\boxL\;.
\label{def:greens_function_rlrw_box}
\end{equation}

These two-point functions are closely-related to one another, as the next lemma illustrates.
Recall that we consider $\torus$ as a module over the commutative ring $T_{L}$, with addition and scalar multiplication defined modulo $L$
in each entry. For each $x\in\doubletorus$, we can then define 
\begin{equation}
  \label{eq:torus to box equivalence relation}
  [x]_L:=\{y\in\doubletorus \;:\; y_i\in\{x_i,-L-x_i\} \text{ for all } i\in[d]\}.
\end{equation}
The partition of $\doubletorus$ into the sets $[x]_L$ defines an equivalence relation on $\doubletorus$, in which the sets $[x]_L$ are the
equivalence classes. The case of $d=1$, corresponding to projecting a cycle onto a path, is illustrated in Fig.~\ref{fig:projection cartoons}.
\begin{figure}
\centering
     \begin{subfigure}{0.455\textwidth}
    \centering
    \begin{tikzpicture}
  \draw (0,0) ellipse (3cm and 1.5cm);
  \filldraw (0,1.5) circle (2pt) node[anchor=north east] {};
  \filldraw (0,-1.5) circle (2pt) node[anchor=north east] {};
  \filldraw (1,1.41) circle (2pt) node[anchor=north east] {};
  \filldraw (1,-1.41) circle (2pt) node[anchor=north east] {};
  \filldraw (2,1.1) circle (2pt) node[anchor=north east] {};
  \filldraw (2,-1.1) circle (2pt) node[anchor=north east] {};
  \filldraw (3,0) circle (2pt) node[anchor=north east] {};
  \filldraw (-1,1.41) circle (2pt) node[anchor=north east] {};
  \filldraw (-1,-1.41) circle (2pt) node[anchor=north east] {};
  \filldraw (-2,1.1) circle (2pt) node[anchor=north east] {};
  \filldraw (-2,-1.1) circle (2pt) node[anchor=north east] {};
  \filldraw (-3,0) circle (2pt) node[anchor=north east] {};

  \draw[dashed] (0,-1.5) -- (0,1.5);
  \draw[dashed] (1,-1.41) -- (1,1.41);
  \draw[dashed] (-1,-1.41) -- (-1,1.41);
  \draw[dashed] (-2,-1.1) -- (-2,1.1);
  \draw[dashed] (2,-1.1) -- (2,1.1);

  \node at (0,-1.9) {$0$};  \node at (-1.1,-1.8) {$-1$}; \node at (1,-1.8) {$1$};
  \node at (-0.1,1.9) {$-6$};  \node at (1,1.8) {$5$}; \node at (-1.1,1.8) {$-5$};
  \node at (2,1.5) {$4$}; \node at (2,-1.5) {$2$};
  \node at (3.3,0) {$3$}; \node at (-3.4,0) {$-3$};
  \node at (-2.1,1.5) {$-4$}; \node at (-2.1,-1.5) {$-2$};
  \end{tikzpicture}
    \caption{\label{subfig:reflection projection}}
   \end{subfigure}
     \begin{subfigure}{0.455\textwidth}
    \centering
    \begin{tikzpicture}
   \draw (7,0) ellipse (3cm and 1.5cm);
  \filldraw (7,1.5) circle (2pt) node[anchor=north east] {};
  \filldraw (7,-1.5) circle (2pt) node[anchor=north east] {};
  \filldraw (7.9,1.41) circle (2pt) node[anchor=north east] {};
  \filldraw (7.9,-1.41) circle (2pt) node[anchor=north east] {};
  \filldraw (8.8,1.2) circle (2pt) node[anchor=north east] {};
  \filldraw (8.8,-1.2) circle (2pt) node[anchor=north east] {};
  \filldraw (9.7,0.65) circle (2pt) node[anchor=north east] {};
  \filldraw (9.7,-0.65) circle (2pt) node[anchor=north east] {};
  \filldraw (6.1,1.41) circle (2pt) node[anchor=north east] {};
  \filldraw (6.1,-1.41) circle (2pt) node[anchor=north east] {};
  \filldraw (5.2,1.2) circle (2pt) node[anchor=north east] {};
  \filldraw (5.2,-1.2) circle (2pt) node[anchor=north east] {};
  \filldraw (4.3,0.65) circle (2pt) node[anchor=north east] {};
  \filldraw (4.3,-0.65) circle (2pt) node[anchor=north east] {};

  \draw[dashed] (7,-1.5) -- (7,1.5);
  \draw[dashed] (7.9,-1.41) -- (7.9,1.41);
  \draw[dashed] (6.1,-1.41) -- (6.1,1.41);
  \draw[dashed] (8.8,-1.1) -- (8.8,1.1);
  \draw[dashed] (9.7,-0.6) -- (9.7,0.6);  
  \draw[dashed] (5.2,-1.1) -- (5.2,1.1);
  \draw[dashed] (4.3,-0.6) -- (4.3,0.6);
  
  \node at (7,-1.9) {$0$}; \node at (7,1.9) {$-7$};
  \node at (7.9,-1.8) {$1$}; \node at (7.9,1.8) {$6$};
  \node at (8.8,1.55) {$5$}; \node at (8.8,-1.55) {$2$};
  \node at (9.7,1.1) {$4$}; \node at (9.7,-1.1) {$3$};
  \node at (6,-1.8) {$-1$}; \node at (6,1.8) {$-6$};  
  \node at (5.1,1.55) {$-5$}; \node at (5.1,-1.55) {$-2$};
  \node at (4.2,1.1) {$-4$}; \node at (4.2,-1.1) {$-3$};
\end{tikzpicture}
    \caption{\label{subfig:holding projection}}
     \end{subfigure}
     \caption{\label{fig:projection cartoons}
       Illustration of the equivalence classes defined by~\eqref{eq:torus to box equivalence relation} with $d=1$.
       (\subref{subfig:reflection projection}) Equivalence classes, $[x]_{L-1}$, on $\TT_{2(L-1)}^1$ with $L=7$.
       (\subref{subfig:holding projection}) Equivalence classes, $[x]_{L}$, on $\TT_{2L}^1$ with $L=7$.
       Note that in both cases the set of equivalence classes are in bijection with $\BB_L^1$.}
\end{figure}

\begin{lemma}
  \label{lemma:two-point function projection identities}
  Let $d,L\in\posint$ and let $x\in\boxL$. Then:
  \begin{enumerate}[label=(\roman*)]    
  \item\label{projection:infinite to torus} For any $L\ge3$
  $$
  g_{\rP,\,L,\,\sN}(x) = \sum_{z\in\ZZ^d} g_{\sN}(x+Lz)
  $$
  \item\label{projection:torus to reflecting} For any odd $L\ge3$
  $$
  g_{\rR,\,L,\,\sN}(x) = \sum_{x'\in[x]_{L-1}} g_{\rP,\,2(L-1),\,\sN}(x')
  $$
  \item\label{projection:torus to holding} For any odd $L\ge3$
  $$
  g_{\rH,\,L,\,\sN}(x) = \sum_{x'\in[x]_L} g_{\rP,\,2L,\,\sN}(x')
  $$
  \end{enumerate}
\end{lemma}
The proof of Lemma~\ref{lemma:two-point function projection identities}, which is based on Markov chain projection arguments, is discussed
in Section~\ref{sec:proof of projection lemma}. 

\subsection{Main results for RLRW two-point functions}
\label{subsec:two-point function propositions}
We now state our main results for the asymptotic behaviour of RLRW two-point functions on $\boxL$, for periodic, reflecting and holding
boundary conditions. We defer proof of these results to Sections~\ref{sec:proof fast decay} and~\ref{sec:proof plateau}. We provide
numerical evidence of the connection of these results to the Ising and SAW models in Section~\ref{subsec:universal two-point functions}. 

The RLRW model on $\boxL$ is most easily understood by relating it to the RLRW on the infinite lattice. 
The asymptotic behaviour of the latter was studied in detail in~\cite{DengGaroniGrimmZhou2022}. See also~\cite{MichtaSlade2022}.

Suppose $\sN_L$ is chosen so that its typical scale $a_L$ grows with $L$. One would expect the behaviour of $g_{\ast,L,\sN_L}(x)$ to differ
qualitatively depending on whether or not $a_L$ grows fast enough that the RLRW can explore distances from the origin of order $L$,
so that the presence of the boundary can be felt. We therefore present two separate results relating $g_{\ast,L,\sN_L}$ to $g_{\sN_L}$,
depending on the asymptotics of $\sN_L$.

Let $\Delta$ denote the standard degenerate distribution function, i.e. $\Delta$ is the indicator function for $[0,\infty)$. 
\begin{proposition}
  \label{theorem:two-point function fast decay}
  Consider a sequence of $\naturals$-valued random variables $\sN_L$, for which there exists a sequence $a_L>0$ satisfying:
  \begin{enumerate}
  \item $a_L\to\infty$.
  \item\label{small lambda a_L bound} $a_L=O(L^\lambda)$ for some $\lambda <2$.
  \item\label{MGF bound assumption} There exists $r,C>0$ such that $\EE(\e^{r\sN_L/a_L})\le C$ for all $L$.
  \item There exists a distribution function $G\neq\Delta$ such that $\PP(\sN_L/a_L\le \cdot)\implies G$.
  \end{enumerate}
Fix $d\ge 3$, and let $(x_L)_{L\ge3}$ be a sequence in $\ZZ^d$ satisfying $x_L\in\boxL$ and $|x_L|/\sqrt{a_L}\to \xi\in[0,\infty)$. 
Then, with $\ast$ denoting P, R or H, as $L\to\infty$ we have:
$$g_{\ast,\,L,\,\sN_L}(x_L) = g_{\sN_L}(x_L)[1+o(1)]$$
\end{proposition}

The assumptions on $\sN_L$ given in Proposition~\ref{theorem:two-point function fast decay} imply that typical $\sN_L$ will have length of
order $a_L=O(L^\lambda)$ with $\lambda<2$. Since a SRW walk of length $N$ typically explores distances from the origin of order $\sqrt{N}$,
it then follows that a typical such RLRW will explore distances of order $o(L)$ from the origin, and will therefore be too short to
feel the boundary. It is therefore unsurprising that the finite-box and infinite lattice two-point functions are asymptotic on such a
scale, for any of the three choices of boundary conditions studied. The spatial scales probed by Proposition~\ref{theorem:two-point function
fast decay} correspond to distances of the order of $\sqrt{a_L}$, where $a_L$ is the typical scale of $\sN_L$. Under the assumptions of
Proposition~\ref{theorem:two-point function fast decay}, it follows from Proposition~\ref{theorem:two-point function fast decay}
and~\cite[Proposition 3.1]{DengGaroniGrimmZhou2022} that
\begin{equation}
  \label{eq:infinite lattice fast decay}
  \lim_{L\rightarrow \infty} \|x_L\|^{d-2} g_{\ast,L,\sN_L}(x_L)  =
  \frac{d}{2\pi^{d/2}}\int_{0}^\infty s^{d/2 - 2}\e^{-s}\,\left[1-G\left(\frac{d}{2}\frac{\xi^2}{s}\right)\right] \diff s
\end{equation}
We note that, in particular, if $G$ corresponds to the distribution function of a mean-$1$ exponential random variable, then we have
\begin{equation}
  \label{eq:fast decay RLRW two-point function}
  \lim_{L\rightarrow \infty} \|x_L\|^{d-2} g_{\ast,L,\sN_L}(x_L)  =   
  \sE(\xi)
 \end{equation}
 with
 \begin{equation}
   \begin{split}
     \sE(\xi)&:= \frac{d}{2\pi^{d/2}}\int_{0}^\infty s^{d/2 - 2}\exp\left(-s-\frac{d}{2}\frac{\xi^2}{s}\right)\, \diff s\\
     &\phantom{:}=
     \frac{2}{\pi^{d/2}}\,\left(\frac{d}{2}\right)^{\frac{d}{4}+\frac{1}{2}}
     \, \xi^{\frac{d}{2}-1} \, K_{\frac{d}{2}-1}\,\left(\sqrt{2d} \xi \right)\\
     &\phantom{:}\sim
     \frac{d^{d+1}{4}}{2^{\frac{d+1}{4}}\,\pi^{\frac{d-1}{2}}}\,\xi^{\frac{d-3}{2}}\,\e^{-\sqrt{2d}\,\xi}, \qquad \xi\to\infty
     \label{eq:Continuum Green function ansatz}
     \end{split}
   \end{equation}
and where $K_\nu(\cdot)$ denotes the modified Bessel function of the second-kind~\cite{GradshteynRyzhik2007}. As discussed 
in~\cite{MichtaSlade2022}, $\sE$ is intimately related to the Green function of the continuum Laplacian on $\RR^d$.
As elaborated on in Section~\ref{sec:numerical results}, the exponential choice for $G$ here is motivated by the behaviour of the self-avoiding walk on the
complete graph.

We now turn our attention to the case that $\EE(\sN_L)=\Omega(L^2)$.
\begin{proposition}
\label{theorem:two-point function plateau}
Fix $d\ge 3$, and let $(x_L)_{L\ge3}$ be a sequence in $\ZZ^d$ satisfying $x_L\in\boxL$. Let $\left(\sN_L\right)_{L\in\posint}$ be a
sequence of $\naturals$-valued random variables. Let $\ast$ denote $\rP$, $\rR$ or $\rH$. Then as $L\to\infty$ we have: 
\begin{enumerate}[label=(\roman*)]
\item\label{eq:plateau theorem weak} If $\EE(\sN_L) = \Omega(L^2)$, then
$$
g_{\ast,\,L,\,\sN_L}(x_L) - g_{\sN_L}(x_L) \asymp \EE(\sN_L)/L^{d}
$$
\item\label{eq:plateau theorem sharp} If $\EE(\sN_L) = \omega(L^2)$, then 
$$
g_{\ast,\,L,\,\sN_L}(x_L) - g_{\sN_L}(x_L) \sim \EE(\sN_L)/L^{d}
$$
\end{enumerate}
\end{proposition}
A similar result is given in~\cite[Theorem 1.4]{Slade2023} for the case of a geometric walk length distribution, giving upper and lower bounds
for the difference between the torus and infinite lattice two-point functions in terms of the susceptibility, but without control of
the constants, and with a lower bound that is weaker by a logarithmic factor when $d=4$. 

Suppose that $\sN_L$ is such that $\sN_L/\EE(\sN_L)$ converges in distribution, and that the limiting distribution function is continuous at
the origin. E.g., this occurs if $\sN_L=|\sK_{L^d}|$ in either the critical window or the high-temperature scaling window. Now suppose that
$\EE(\sN_L)\asymp L^{\lambda}$ with $\lambda>2$. It then follows\footnote{Although the statement of~\cite[Proposition
    3.1]{DengGaroniGrimmZhou2022} specifies $\xi>0$, its proof also holds when $\xi=0$.} from~\cite[Proposition
  3.1]{DengGaroniGrimmZhou2022} that for any sequence $(x_L)_{L\ge3}\subset\ZZ^d$ satisfying $x_L\in\boxL$ and $|x_L|\to\infty$ we have 
\begin{equation}
  \label{eq:degenerate limit of infinite lattice two-point function}
  \lim_{L\to\infty} |x_L|^{d-2}\,g_{\sN_L}(x_L) = \frac{d}{2\pi^{d/2}}\,\Gamma(d/2-1).
\end{equation}
Therefore, the exponential decay displayed by~\eqref{eq:fast decay RLRW two-point function} for $\lambda<2$ cannot be observed when $\lambda>2$. Consequently,
Proposition~\ref{theorem:two-point function plateau} implies that $g_{\ast,L,\sN_L}(x_L)$ decays as a power-law for
$|x_L|=o(L^{(d-\lambda)/(d-2)})$, but is then dominated by a term of order $L^{\lambda-d}$ for $|x_L|=\omega(L^{(d-\lambda)/(d-2)})$. We
note that the scale $L^{(d-\lambda)/(d-2)}$ is $o(L)$, and therefore realisable inside $\boxL$, iff $\lambda>2$. To probe the crossover from
power-law to plateau behaviour we choose $x_L\in\boxL$ such that $|x_L|=\varphi\,\left(L^d/\EE(\sN_L)\right)^{1/(d-2)}$ for
$\varphi\in(0,\infty)$, to obtain
\begin{equation}
  \lim_{L\to\infty} |x_L|^{d-2}\,g_{\ast,L,\sN_L}(x_L) = \frac{d}{2\pi^{d/2}}\,\Gamma(d/2-1) + \varphi^{d-2}\;.
\end{equation}

\section{Numerical results}
\label{sec:numerical results}
\subsection{Universal walk length distributions}
\label{subsec:universal walk length distributions}
\begin{figure}
  \centering
   \begin{subfigure}{0.425\textwidth}
    \centering
    \includegraphics[width=\textwidth]{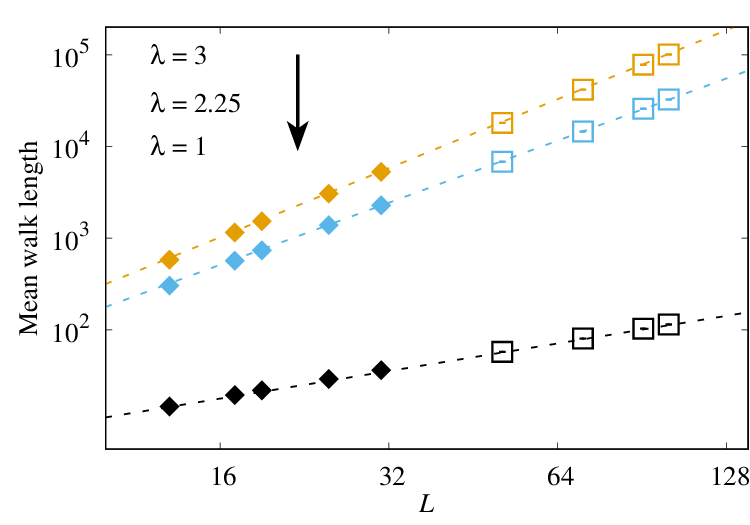}
    \caption{\label{subfig:walk length mean}}
   \end{subfigure}
   \begin{subfigure}{0.425\textwidth}
    \centering
    \includegraphics[width=\textwidth]{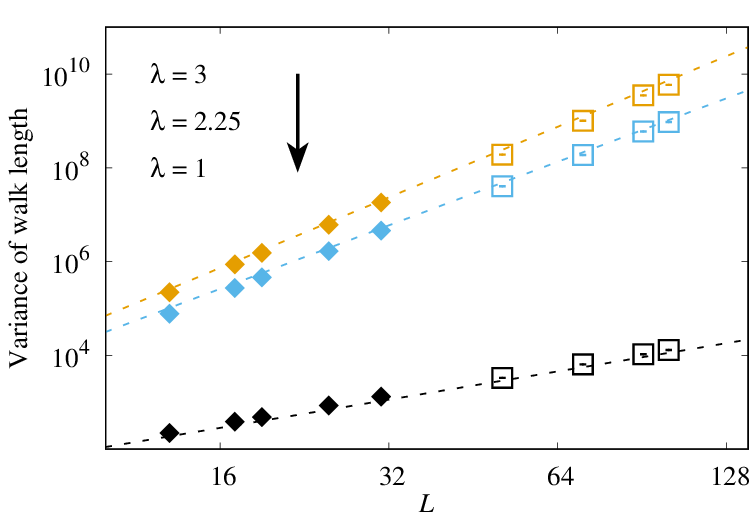}
    \caption{\label{subfig:walk length variance}}
   \end{subfigure}
   \caption{Simulated mean and variance of the Ising (rhombi) and SAW (squares) walk lengths on five-dimensional tori.
     (\subref{subfig:walk length mean}) Simulated $\EE(|\sS|)$ and $\EE(|\sT|)$ at fugacity $z_L = z_c(1-L^{-\lambda})$ with
     $\lambda=1,9/4,3$, on a log-log scale. The dashed curves passing through the $\lambda=1,9/4$ have slope $\lambda$, while the curve
     passing through the $\lambda=3$ data has slope $d/2$.
     (\subref{subfig:walk length variance}) Simulated $\var(|\sS|)$ and $\var(|\sT|)$ at fugacity $z_L = z_c(1-L^{-\lambda})$ with $\lambda=1,9/4,3$.
     The dashed curves passing through the $\lambda=1,9/4$ have slope $2\lambda$, while the curve passing through the $\lambda=3$ data has slope $d$.
   }
\label{fig:walk length mean and variance on torus}
\end{figure}

Figure~\ref{subfig:walk length mean} shows the simulated results for $\EE(|\sS|)$ and $\EE(|\sT|)$ at fugacity $z=\zc(1-L^{-\lambda})$ with
$\lambda=1,9/4,3$. If the boundary of the critical windows for the SAW and Ising model on $\torus$ occur at the square root of the volume,
as occurs for the complete graph SAW, then for $d=5$ the value $\lambda=3>d/2$ should lie inside the critical window while $\lambda=1,9/4$
should lie outside the critical window. Figure~\ref{subfig:walk length mean} is clearly consistent with the conjecture that $\EE(|\sS|)$ and
$\EE(|\sT|)$ scale like $L^\lambda$ when $\lambda<d/2$, but like $L^{d/2}$ for $\lambda>d/2$. Likewise, the results in
Figure~\ref{subfig:walk length variance} are consistent with the conjecture that $\var(|\sS|)$ and $\var(|\sT|)$ scale like $L^{2\lambda}$
when $\lambda<d/2$, but like $L^{d}$ for $\lambda>d/2$. This behaviour is precisely analogous to complete graph SAW behaviour shown
in~\eqref{eq:Kn SAW cumulants - high temperature scaling window} and~\eqref{eq:Kn SAW cumulants - critical window}. 

\begin{figure}
  \centering
     \begin{subfigure}{0.455\textwidth}
    \centering
    \includegraphics[width=\textwidth]{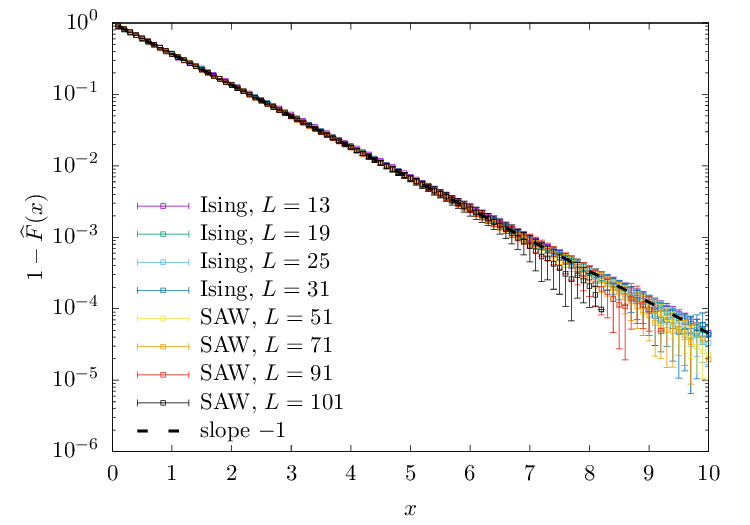}
    \caption{\label{subfig:walk length distribution lambda = 1}}
   \end{subfigure}
     \begin{subfigure}{0.455\textwidth}
    \centering
    \includegraphics[width=\textwidth]{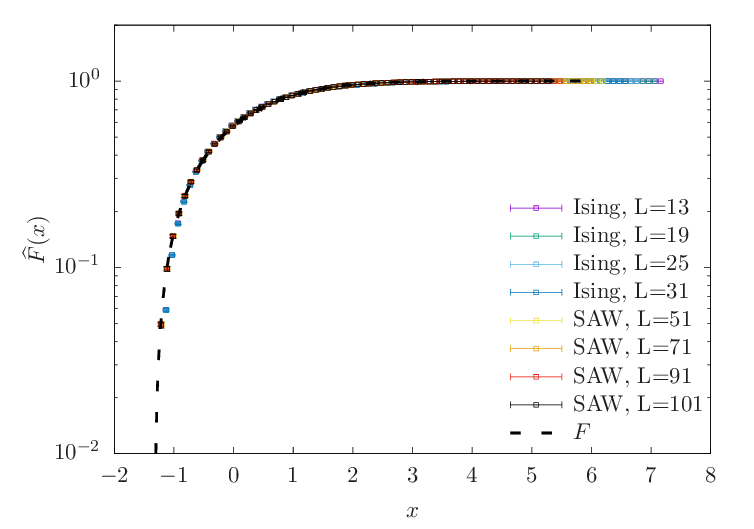}
    \caption{\label{subfig:walk length distribution lambda = 3}}
     \end{subfigure}
     \caption{(\subref{subfig:walk length distribution lambda = 1})
       Tail of the simulated distribution function, $\widehat{F}$, of the rescaled SAW length, $|\sS|/\EE(|\sS|)$, and rescaled Ising
       walk length, $|\sT|/\EE(|\sT|)$, on five-dimensional tori at fugacity $z_L = z_c(1-L^{-\lambda})$ with $\lambda = 1$. The dashed
       curve is $\e^{-x}$, the tail of the mean-1 exponential distribution.
       (\subref{subfig:walk length distribution lambda = 3})
       Simulated distribution function, $\widehat{F}$, of the standardised SAW length, $(|\sS| - \EE(|\sS|))/\sqrt{\var(|\sS|)}$, and
       standardised Ising walk length, $(|\sT| - \EE(|\sT|))/\sqrt{\var(|\sT|)}$, on five-dimensional tori, at fugacity $z_L =
       z_c(1-L^{-\lambda})$ and $\lambda=3$. The dashed curve corresponds to the standardised half-normal distribution function, $F$, given
       in~\eqref{eq:def of standardised half normal}. 
     }
     \label{fig:walk length distributions on torus}
\end{figure}

Similarly, Figure~\ref{fig:walk length distributions on torus} illustrates the appropriately scaled/standardised distribution functions of $|\sS|$
and $|\sT|$ for $\lambda=1$ and $\lambda=3$. Figure~\ref{subfig:walk length distribution lambda = 1} strongly suggests for $\lambda=1$ that
$|\sS|/\EE(|\sS|)$ and $|\sT|/\EE(|\sT|)$ converge in distribution to a mean-1 exponential random variable, precisely as stated
in~\eqref{eq:scaling window Kn limit theorem} for the complete graph SAW in the high-temperature scaling window. Likewise,
Figure~\ref{subfig:walk length distribution lambda = 3} strongly suggests for $\lambda=3$ that $(|\sS|-\EE(|\sS|))/\sqrt{\var(|\sS|)}$ and
$(|\sT|-\EE(|\sT|))/\sqrt{\var(|\sT|)}$ converge in distribution to a standardised half-normal distribution, precisely as stated
in~\eqref{eq:critical window Kn limit theorem} for the complete graph SAW in the critical window.

We note that the same critical window behaviour for the Ising and SAW mean, variance and limit distribution were observed at the estimated
infinite-volume critical fugacity, $\zc$, on 5-dimensional tori in~\cite{DengGaroniGrimmZhou2022}. One can formally view this case as
$\lambda=+\infty$.

To support the claim that the boundary of the critical window lies at $d/2$, Table~\ref{table:exponents} provides estimates of the scaling
exponent $\mu$ obtained by fitting $\EE(|\sS|)$ and $\EE(|\sT|)$ to an ansatz $a+b L^\mu$, for various values of $\lambda$. As expected from
the complete graph SAW results, we indeed observe that $\mu=\lambda$ for each $\lambda\le d/2$, but $\mu=d/2$ for all $\lambda\ge d/2$.
\begin{table}[h]
\centering
\caption{Estimated $\mu$ values for $\EE(|\sS|)$ and $\EE(|\sT|)$ on $\torus$ with $d=5$ at fugacity $z_L=\zc(1-L^{\lambda})$ at
  various values of $\lambda$.}
\label{table:exponents}
\begin{tabular}{|c|c|c|}
\hline 
$\lambda$ & $\EE(|\sT|)$ & $\EE(|\sS|)$ \\
\hline 
$1$   & 1.00(1) & 0.998(2) \\
\hline 
$3/2$ & 1.53(5) & 1.499(2) \\
\hline 
$2$   & 2.01(9) & 2.01(1) \\
\hline 
$5/2$ & 2.50(5) & 2.46(4) \\
\hline 
$3$   & 2.51(2) &  2.5(1) \\
\hline 
\end{tabular} 
\end{table}

Based on the above observations, we conjecture that the following holds for any $d\ge5$ and $z_L=\zc - a\, L^{-\lambda}$. 
If $\lambda<d/2$ and $a>0$ then 
\begin{equation}
  |\sS|/\EE(|\sS|),\quad |\sT|/\EE(|\sT|) \implies Y
  \label{eq:high T SAW and Ising limit conjecture}
\end{equation}
and there exist constants $\sA_{\sS,d,\lambda,a},\sA_{\sT,d,\lambda,a},\sB_{\sS,d,\lambda,a},\sB_{\sT,d,\lambda,a}>0$ such that
\begin{equation}
  \EE(|\sS|) \sim \sA_{\sS,d,\lambda,a}\,L^{\lambda}\;, \qquad   \EE(|\sT|) \sim \sA_{\sT,d,\lambda,a}\,L^{\lambda}\;, \qquad 
  \label{eq:high T SAW and Ising mean conjecture}
\end{equation}
and
\begin{equation}
  \var(|\sS|) \sim \sB_{\sS,d,\lambda,a}\,L^{2\lambda}\;, \qquad   \var(|\sT|) \sim \sB_{\sT,d,\lambda,a}\,L^{2\lambda}\;.
  \label{eq:high T SAW and Ising var conjecture}  
\end{equation}
While if $\lambda>d/2$, then for any $a\in\RR$
\begin{equation}
  (|\sS|-\EE(|\sS|))/\sqrt{\var(|\sS|)},\quad (|\sT|-\EE(|\sT|))/\sqrt{\var(|\sT|)} \implies \frac{|X|-\EE|X|}{\sqrt{\var(|X|)}}
  \label{eq:critical window SAW and Ising mean conjecture}    
\end{equation}
and there exist constants $\sA_{\sS,d},\sA_{\sT,d},\sB_{\sS,d},\sB_{\sT,d}>0$ such that
\begin{equation}
  \EE(|\sS|) \sim \sA_{\sS,d}\,L^{d/2}\;, \qquad   \EE(|\sT|) \sim \sA_{\sT,d}\,L^{d/2}\;, \qquad 
\end{equation}
and
\begin{equation}
  \var(|\sS|) \sim \sB_{\sS,d}\,L^d\;, \qquad   \var(|\sT|) \sim \sB_{\sT,d}\,L^{d}\;.
\end{equation}

\subsection{Universal two-point functions}
\label{subsec:universal two-point functions}
We now provide numerical evidence that in both the high-temperature scaling window and the critical window, the two-point functions of the
SAW and Ising model display the same asymptotic behaviour as does a RLRW whose walk length distribution is chosen to be that of a corresponding
complete-graph SAW.

\begin{figure}[ht!]
  \centering
   \begin{subfigure}{0.425\textwidth}
    \centering
    \includegraphics[width=\textwidth]{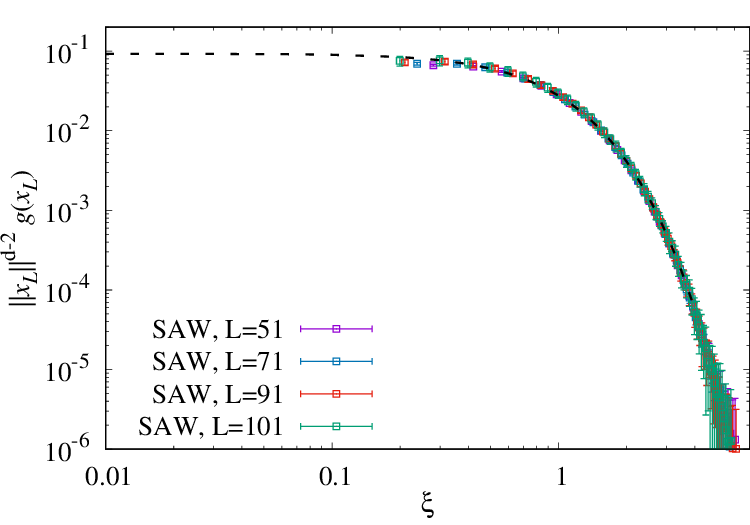}
    \caption{\label{subfig:g SAW lambda=1}}
   \end{subfigure}
   \begin{subfigure}{0.425\textwidth}
    \centering
    \includegraphics[width=\textwidth]{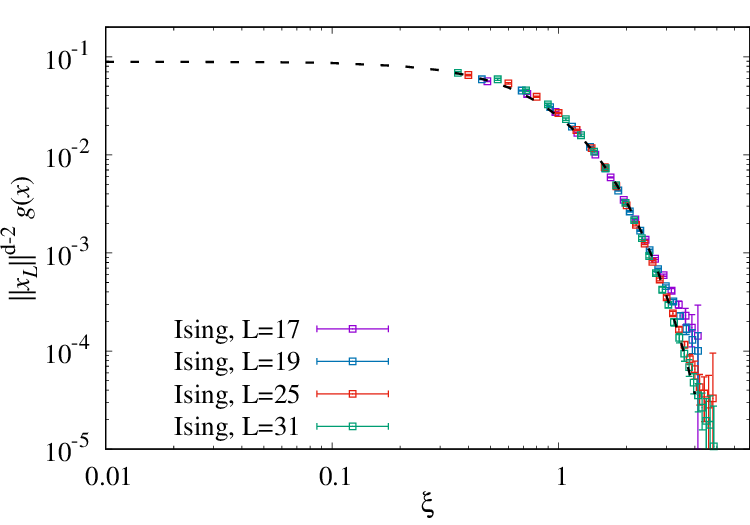}
    \caption{\label{subfig:g Ising lambda=1}}
  \end{subfigure}
   \begin{subfigure}{0.425\textwidth}
    \centering
    \includegraphics[width=\textwidth]{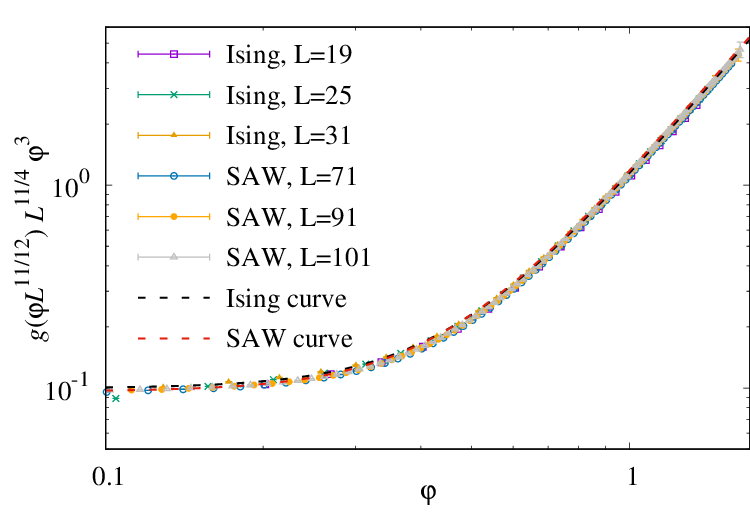}
    \caption{\label{subfig:g SAW+Ising lambda=9/4}}
  \end{subfigure}
   \begin{subfigure}{0.425\textwidth}
    \centering
    \includegraphics[width=\textwidth]{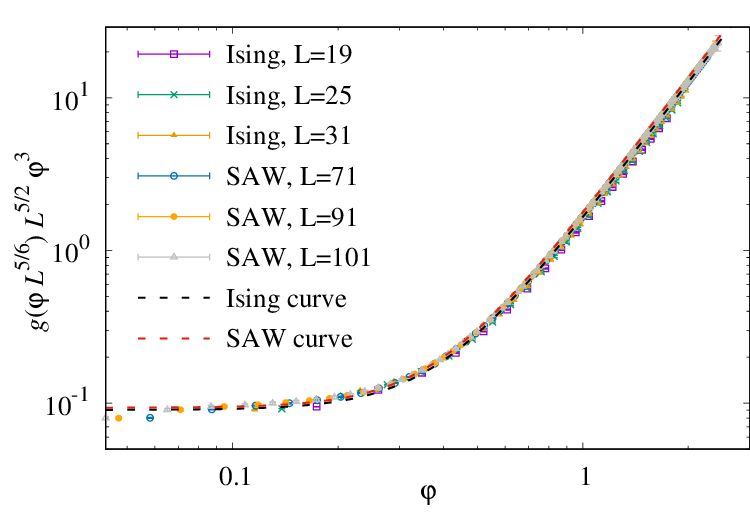}
    \caption{\label{subfig:g SAW+Ising lambda=3}}
  \end{subfigure}
   \caption{(\subref{subfig:g SAW lambda=1}) Two-point functions on five-dimensional tori of SAW at fugacity $z_L=z_c(1-L^{-\lambda})$ and
     $\lambda=1$. The dashed curve corresponds to the ansatz~\eqref{eq:SAW and Ising high T g conjecture} with~\eqref{eq:Continuum Green
       function ansatz}, with constants $\alpha,\gamma$ set to the values described in the text, with $A_{\sS,d}$ estimated via simulation. 
     (\subref{subfig:g Ising lambda=1})
     Analogous plot to (\subref{subfig:g SAW lambda=1}), for Ising case.
     (\subref{subfig:g SAW+Ising lambda=9/4})
     Two-point functions on five-dimensional tori of SAW at fugacity $z_L=z_c(1-L^{-\lambda})$ and $\lambda=9/4$. The dashed curve corresponds
     to the plateau ansatz~\eqref{eq:Ising and SAW plateau}, with constants $\alpha,\gamma$ set to the values described in the text,
     with $B_{\sS,d}$ and $B_{\sT,d}$ estimated via simulation. 
     (\subref{subfig:g SAW+Ising lambda=3}) Analogous plot to (\subref{subfig:g SAW+Ising lambda=9/4}), with $\lambda=3$.
}
\label{fig:unwrapped_two-point_function} 
\end{figure}

We first consider the high-temperature scaling window with $\lambda<2$. Assuming the validity of the conjectures on $\sS$ and $\sT$ outlined
in~\eqref{eq:high T SAW and Ising limit conjecture} and~\eqref{eq:high T SAW and Ising mean conjecture},
it follows from standard convergence of types arguments (see e.g.~\cite[pp. 193]{Billingsley1995}) that for all $y\in\RR$
\begin{equation}
  \lim_{L\to\infty}\PP\left(\frac{|\sS|}{L^{\lambda}}\le y\right)= \left(1 - \e^{-y/\sA_{\sS,d,\lambda,a}}\right)\ind(y\ge0).
  \label{eq:conjectured limit of SAW distribution function}
\end{equation}
Now let $\sN_L=|\sS_L|$, $a_L=L^{\lambda}$ and for fixed $\xi\in(0,\infty)$ let $x_L$ satisfy $|x_L|= L^{\lambda/2}\xi$. Assuming the
validity of~\eqref{eq:conjectured limit of SAW distribution function}, and that the assumptions of Proposition~\ref{theorem:two-point
  function fast decay} hold for this choice of $\sN_L$, it follows from\eqref{eq:fast decay RLRW two-point function} that as $L\to\infty$
\begin{equation}
  |x_L|^{d-2}\,g_{\rP,L,\sN_L}(x_L) \sim \sE(\xi/\sqrt{\sA_{\sS,d,\lambda,a}}).
  \label{eq:unwrapped RLRW g with SAW length}
\end{equation}
Universality then makes it natural to conjecture that the asymptotics of $|x_L|^{d-2}\,g(x_L)$ for the SAW and Ising model on
the torus should be given by
\begin{equation}
  |x_L|^{d-2}\,g(x_L) \sim \alpha \,\sE(\gamma\, \xi)
  \label{eq:SAW and Ising high T g conjecture}
\end{equation}
for suitable values of the model-dependent constants, $\alpha,\gamma$, depending $d$, $\lambda$, and $a$. Figures~\ref{subfig:g SAW
  lambda=1} and~\ref{subfig:g Ising lambda=1} provide strong evidence in favour of these conjectures. In Figure~\ref{subfig:g SAW lambda=1},
the constants for SAW are set to $\alpha=0.75$, $\gamma=0.83/\sqrt{A_{\sS,d,\lambda,a}}$, while in Figure~\ref{subfig:g Ising lambda=1} the constants for the
Ising model are set to $\alpha = 0.75$, $\gamma=0.87/\sqrt{A_{\sT,d,\lambda,a}}$, where $\sA_{\sS,d,\lambda,a}$ and $\sA_{\sT,d,\lambda,a}$
were estimated by fitting the mean walk length; cf.~\eqref{eq:high T SAW and Ising mean conjecture}.

We now consider $\lambda>2$. Assuming the validity of the conjecture~\eqref{eq:high T SAW and Ising mean conjecture}, the discussion in
Section~\ref{subsec:two-point function propositions} suggests that we should observe a plateau in this case. Moreover, we expect the order
of the plateau to be $L^{\lambda-d}$ for $\lambda<d/2$ and $L^{d/2}$ for $\lambda>d/2$. More concretely, suppose $2<\lambda<d/2$ and let
$\sN_L=|\sS_L|$, and for fixed $\varphi\in(0,\infty)$ let $x_L$ satisfy $|x_L|= L^{(d-\lambda)/(d-2)}\,\varphi$. Assuming the validity of
the conjecture~\eqref{eq:high T SAW and Ising mean conjecture}, it follows from Proposition~\ref{theorem:two-point function plateau}
and~\eqref{eq:degenerate limit of infinite lattice two-point function} that 
\begin{equation}
  |x_L|^{d-2}\,g_{\rP,L,\sN_L}(x_L) \sim \frac{d}{2\pi^{d/2}}\Gamma(d/2-1)+ \sA_{\sS,d,\lambda,a} \,\varphi^{d-2}\;.
\end{equation}
Universality then makes it natural to conjecture that for both the SAW and Ising model on the torus
\begin{equation}
  |x_L|^{d-2}\,g(x_L) \sim \alpha\,\frac{d}{2\pi^{d/2}}\Gamma(d/2-1)+ \gamma\,\varphi^{d-2}
  \label{eq:Ising and SAW plateau}
\end{equation}
for suitable model-dependent parameters $\alpha,\gamma$, depending on $d,\lambda$ and $a$. Figure~\ref{subfig:g SAW+Ising lambda=9/4}
plots the SAW and Ising cases on tori with $d=5$, $\lambda=9/4$ and $a=\zc$, with the constants set to $\alpha=0.76$ and $\gamma=1.08\sA_{\sS,d,\lambda,a}$ for SAW and
$\alpha=0.79$ and $\gamma=1.05\sA_{\sT,d,\lambda,a}$ for the Ising model.

Finally, we consider the case $\lambda>d/2$, which lies inside the critical window. Let $\sN_L=|\sS_L|$, and for fixed
$\varphi\in(0,\infty)$ let $x_L$ satisfy $|x_L|= L^{(d/2)/(d-2)}\,\varphi$. Then assuming the validity of the conjecture~\eqref{eq:high T
  SAW and Ising mean conjecture}, it follows that~\eqref{eq:Ising and SAW plateau} should hold for suitable model-dependent parameters
$\alpha,\gamma$, depending on $d$. Figure~\ref{subfig:g SAW+Ising lambda=3} plots the SAW and Ising cases on tori with $d=5$, $\lambda=3$
and $a=\zc$, with the constants set to $\alpha=0.74$ and $\gamma=1.71\sA_{\sS,d}$ for SAW, and $\alpha=0.71$ and $\gamma=1.59\sA_{\sT,d}$ for the Ising model.

\section{Proof of Proposition~\ref{theorem:two-point function fast decay}}
\label{sec:proof fast decay}
Let $(S_n)_{n=0}^{\infty}$ be a simple random walk on $\ZZ^d$, starting from the origin, and let
\begin{equation}
p_n(z):=\PP(S_n=z), \qquad z\in\ZZ^d.
\end{equation}
We say that $n\in\naturals$ and $z\in\ZZ^d$ have the same parity, and write $n\leftrightarrow z$, iff $n + |z|_1$ is even, where
$|\cdot|_1$ denotes the $\ell^1$ norm on $\RR^d$. Clearly, $p_n(z) = 0$ if $n\nleftrightarrow
z$. Rearranging~\eqref{def:greens_function_rlrw} we obtain
\begin{equation}
  g_{\sN}(z) = \sum_{n=0}^\infty \,p_n(z)\;\PP(\sN\ge n)\;. 
\label{eq:concrete expression for g_Z}
\end{equation}

The proof of Proposition~\ref{theorem:two-point function fast decay} will utilise the following three lemmas, whose proofs are deferred until
the end of this section.
\begin{lemma}
  \label{lemma:exponential upper bound on p_n}
  Let $d\in\posint$. Then for all $n\in\posint$ and all $x\in\ZZ^d$
  $$
  p_n(x)\ll \e^{-|x|/\sqrt{n}}\;.
  $$
\end{lemma}

\begin{lemma}
\label{lemma:remove x in (x+zL)^2}
Let $L$ and $d$ be positive integers. For all $z\in\ZZ^d\setminus0$ and all $x\in [-L/2,L/2]^d$
\begin{equation}
  \label{eq:generic norm bound}
  \frac{1}{2}\,|z|\,L \le |x + z L| \le 2\sqrt{d} \,|z|\,L\;.
\end{equation}
\end{lemma}

\begin{lemma}
  \label{lemma:lower bound on infinite lattice two-point function}
  Consider $\naturals$-valued random variables $\sN_L$ for which there exists $a_L>0$ satisfying $a_L\to\infty$ and a distribution function
  $G\neq\Delta$ such that $\PP(\sN_L/a_L\le \cdot)\implies G$. Let $(x_L)_{L\ge3}$ be a sequence in $\ZZ^d$ satisfying $x_L\in\boxL$ and
  $|x_L|/\sqrt{a_L}\to\xi\in[0,\infty)$. Then
    $$
    g_{\sN_L}(x_L)\gg_\xi \,a_L^{-d/2}\;.
    $$
\end{lemma}

\begin{proof}[Proof of Proposition~\ref{theorem:two-point function fast decay}]
Let $(x_L)_{L\ge3}$ be a sequence in $\ZZ^d$ satisfying $x_L\in\boxL$. In all that follows, any reference to reflecting or holding boundary
conditions on $\boxL$ assumes $L$ is odd.

First note that Lemma~\ref{lemma:two-point function projection identities} implies
  \begin{equation}
    \label{eq:lower bound on g ratio in small lambda}
    \frac{g_{\ast,\,L,\,\sN_L}(x_L)}{g_{\sN_L}(x_L)} - 1> 0,
  \end{equation}
  with $\ast=P,R,H$. It therefore suffices to show that the left-hand side of~\eqref{eq:lower bound on g ratio in small lambda} is $o(1)$ as
  $L\to\infty$.

  We first consider the case of periodic boundary conditions. Part~\ref{projection:infinite to torus} of Lemma~\ref{lemma:two-point
    function projection identities} implies
  \begin{equation}
    \label{eq:simple torus decomposition}
    g_{\rP,\,L,\,\sN_L}(x_L)= g_{\sN_L}(x_L) + \sum_{z\in\ZZ^d\setminus 0}\,g_{\sN_L}(x_L + L\,z)
  \end{equation}
  and since $p_0(z)=0$ for $z\neq 0$, Eq.~\eqref{eq:concrete expression for g_Z} gives
  \begin{equation}
    \label{eq:concrete torus decomposition}
    g_{\rP,\,L,\,\sN_L}(x_L) =
    g_{\sN_L}(x_L)+ \sum_{z\in\ZZ^d\setminus 0}\,\sum_{n=1}^\infty p_n(x_L+L\,z)\;\PP(\sN_L\ge n)\;.
\end{equation}
We begin by showing that the second term on the right-hand side of~\eqref{eq:concrete torus decomposition} is exponentially small in $L$.
First consider the large $n$ terms. Combining assumption~\eqref{MGF bound assumption} with the Chernoff bound
implies that $\PP(\sN_L\ge n)\le C \e^{-r\, n/a_L}$. Letting $\zeta_{L}:=\lfloor L^{1+\lambda/2}\rfloor$, and recalling
assumption~\eqref{small lambda a_L bound}, it then follows that for any $S\subseteq\ZZ^d$ we have
\begin{equation}
  \label{large n bound in small lambda prop}
  \sum_{y\in S} \,\sum_{n=\zeta_{L}}^\infty p_n(y)\; \PP(\sN_L\ge n)
  \ll \sum_{n=\lfloor L^{1+\lambda/2}\rfloor}^\infty \e^{-rn/a_L}
  \ll \exp(-\kappa\, L^{\frac{2-\lambda}{2}})
\end{equation}
for some $\kappa>0$ which depends on $d$ and on the specific sequence of distributions corresponding to $(\sN_L)$.

Now consider the small $n$ terms. Fix $c>0$ and let $y\in\ZZ^d$ satisfy $|y|\ge c\,L$. Lemma~\ref{lemma:exponential upper bound on
  p_n} implies
\begin{equation}
  \label{small n bound in small lambda prop}
  \begin{split}
  \sum_{n=1}^{\zeta_{L}-1} p_n(y) &\ll \sum_{n=1}^{\zeta_{L}} \e^{-|y|/\sqrt{n}} \\
  &\ll \sum_{n=1}^{\zeta_{L}} \e^{-c\,L/ \sqrt{n}}\\
  &\ll L^{1+\lambda/2} \exp\left(-c\,L^{\frac{2-\lambda}{4}}\right)\\
  \end{split}
\end{equation}
But Lemma~\ref{lemma:remove x in (x+zL)^2} implies $|x_L+zL|\ge |z| L/2$ for all $z\in\ZZ^d\setminus0$ and $x_L\in\boxL$, and
so~\eqref{small n bound in small lambda prop} implies 
 \begin{equation}
   \begin{split}
   \sum_{z\in\ZZ^d\setminus 0}\, \sum_{n=1}^{\zeta_L-1} p_n(x_L+L\,z)
   &\ll L^{1+\lambda/2}\,\sum_{z\in\ZZ^d\setminus0} \exp\left(-|z|L^{\frac{2-\lambda}{4}}/2\right) \\
   &\ll L^{1+\lambda/2}\,\,\exp\left(-L^{\frac{2-\lambda}{4}}/4\right)\,\sum_{z\in\ZZ^d\setminus0}\exp\left(-|z|/4\right) 
   \end{split}
   \label{small n bound in small lambda prop concrete}
   \end{equation}
Therefore combining \eqref{large n bound in small lambda prop} and~\eqref{small n bound in small lambda prop concrete}
with~\eqref{eq:concrete torus decomposition} shows that 
\begin{equation}
  \label{eq:nonzero z sum exponentially small}
  g_{\rP,\,L,\,\sN_L}(x_L)-g_{\sN_L}(x_L) \ll L^{1+\lambda/2}\, \exp(-L^{\frac{2-\lambda}{4}}/4)\;.
\end{equation}

We now prove an analogous result for reflecting/holding boundaries. From Parts~\ref{projection:torus to reflecting}
and~\ref{projection:torus to holding} of Lemma~\ref{lemma:two-point function projection identities} we have for $\ast=R,H$
\begin{equation}
  \begin{split}
  \label{eq:reflecting/holding to finite decomposition}
  g_{\ast,\,L,\,\sN_L}(x_L)
  &=
  g_{\sN_L}(x_L) + 
  \sum_{y\in[x_L]_{\tL}\setminus x_L}\,\sum_{n=1}^\infty\,p_n(y)\,\PP(\sN_L\ge n) \\
  &\qquad + \sum_{y\in[x_L]_{\tL}}\,\sum_{z\in\ZZ^d\setminus 0}\,\sum_{n=1}^\infty\,p_n(y+2\,\tL\,z)\,\PP(\sN_L\ge n)
  \end{split}
\end{equation}
where $\tL=L$ if $\ast=\rH$, and $\tL=L-1$ if $\ast=\rR$.
The second and third terms on the right-hand side in~\eqref{eq:reflecting/holding to finite decomposition} can be shown to be exponentially
small by arguing analogously to the periodic case. Indeed, applying~\eqref{large n bound in small lambda prop} immediately shows that 
\begin{multline}
  \sum_{y\in[x_L]_{\tL}\setminus x_L}\,\sum_{n=\zeta_L}^\infty\,p_n(y)\,\PP(\sN_L\ge n) \,+\,
  \sum_{y\in[x_L]_{\tL}}\,\sum_{z\in\ZZ^d\setminus 0}\,\sum_{n=\zeta_L}^\infty\,p_n(y+2\,\tL\,z)\,\PP(\sN_L\ge n)\\
  \ll \exp(-\kappa\, L^{\frac{2-\lambda}{2}})\;.
\end{multline}
Now, $[x_L]_{\tL}$ is a subset of $\doubleboxL$ of fixed cardinality, $2^d$, and so arguing analogously to~\eqref{small n bound in small lambda
  prop concrete} implies
\begin{equation}
  \sum_{y\in[x_L]_{\tL}}\,\sum_{z\in\ZZ^d\setminus 0}\,\sum_{n=1}^{\zeta_L-1}\,p_n(y+2\,\tL\,z)\,\PP(\sN_L\ge n)
  \ll L^{1+\lambda/2}\,\exp(-L^{\frac{2-\lambda}{4}}/4)\;.
\end{equation}
Similarly, since $|y|\ge L/4$ for all $y\in[x_L]_{\tL}\setminus x_L$, it follows immediately from~\eqref{small n bound in small lambda prop}
that
\begin{equation}
  \sum_{y\in[x_L]_{\tL}\setminus x_L}\,\sum_{n=1}^{\zeta_L-1}\,p_n(y)\,\PP(\sN_L\ge n)
  \ll L^{1+\lambda/2}\,\exp(-L^{\frac{2-\lambda}{4}}/4)\;.
\end{equation}
Summarising then, we have for $\ast=P,R,H$ that
\begin{equation}
  g_{\ast,\,L,\,\sN_L}(x_L)- g_{\sN_L}(x_L)
  \ll L^{1+\lambda/2}\,\exp(-L^{\frac{2-\lambda}{4}}/4)\;.
\label{eq:small lambda main asymptotic result}
\end{equation}

But Lemma~\ref{lemma:lower bound on infinite lattice two-point function} implies that
$g_{\sN_L}(x_L)\gg_\xi a_L^{-d/2}$, and so it follows from~\eqref{eq:small lambda main asymptotic result} that
for $\ast=P,R,H$ 
\begin{equation}
    \label{decomposing g in small lambda prop}
  0\le \frac{g_{\ast,\,L,\,\sN_L}(x_L)}{g_{\sN_L}(x_L)}-1 \ll_\xi L^{1+(d+1)\lambda/2}\,\exp(-L^{\frac{2-\lambda}{4}}/4)\;.
\end{equation}
\end{proof}

\begin{proof}[Proof of Lemma~\ref{lemma:exponential upper bound on p_n}]
Let $a\ge0$, and for $j\in[d]$ let $S_n^j$ denote the $j$th coordinate of $S_n$. We begin with the elementary observation that if
$(S_n^j)^2<a^2/d$ for all $j\in[d]$ then $\sum_{j\in[d]}(S_n^j)^2<a^2$. It then follows from the union bound that 
\begin{equation}
  \begin{split}
    \PP(|S_n|\ge a) &\le \sum_{j\in[d]} \PP\left[|S_n^j|\ge a/\sqrt{d}\right] \\
    &=2d\,\PP[S_n^1 \ge a/\sqrt{d}] \\
    &\le 2d\, \EE(\e^{\lambda_n S_n^1})\, \e^{-\lambda_n a/\sqrt{d}}
    \end{split}
\end{equation}
for any $\lambda_n>0$, where in the last step we utilised the Chernoff bound. Consequently, for any $x\in\ZZ^d$ we have
\begin{equation}
  p_n(x) \le 2d\, \EE(\e^{\lambda_n S_n^1})\, \e^{-\lambda_n |x|/\sqrt{d}}.
    \label{Chernoff for SRW MGF}
\end{equation}
A simple calculation shows that for all $t\in\RR$
\begin{equation}
  \EE(\e^{t S_n^1}) = \left[1-\frac{1}{d}+\frac{\cosh(t)}{d}\right]^n.
\end{equation}
Therefore, taking $\lambda_n=\sqrt{d/n}$ we have as $n\to\infty$ that 
\begin{equation}
\label{bound on SRW MGF}
  \log\EE(\e^{\sqrt{d} S_n^1/\sqrt{n}}) = \frac{1}{2} + O(n^{-1}).
\end{equation}
The stated result then follows from~\eqref{bound on SRW MGF} and the specialisation of~\eqref{Chernoff for SRW MGF} to
$\lambda_n=\sqrt{d/n}$.
\end{proof}

\begin{proof}[Proof of Lemma~\ref{lemma:remove x in (x+zL)^2}]
  First note that $(a+y)^2\ge y^2/4$ for all $a\in[-1/2,1/2]$ and $y\in\ZZ$.
  It follows that for any $x\in[-L/2,L/2]^d$ and $z\in\ZZ^d\setminus0$
  $$
  \left|\frac{x}{L}+z\right|^2 =\sum_{i=1}^d \left(\frac{x_i}{L}+z_i\right)^2 \ge \sum_{i=1}^d\frac{z_i^2}{4} =\frac{|z|^2}{4},
  $$
  which establishes the lower bound in~\eqref{eq:generic norm bound}. 
  
  Now, since $|x+zL|^2=|x|^2+L^2|z|^2+2L\,x\cdot z$, the Cauchy-Schwarz inequality implies 
  \begin{equation}
    |x+zL|^2 
    \le
    |z|^2\,L^2\,\left(1+\frac{|x|}{L\,|z|}\right)^2    
    \label{eq:main norm inequality}
  \end{equation}
  Since $|x|\le\sqrt{d}\,L/2$, we have
  \begin{equation}
    |x+zL| \le \left(1+\frac{\sqrt{d}}{2}\right)\, |z|\,L
    \label{eq:generic norm upper bound}    
  \end{equation}
  This establishes the upper bound in~\eqref{eq:generic norm bound}.
\end{proof}

\begin{proof}[Proof of Lemma~\ref{lemma:lower bound on infinite lattice two-point function}]
  Since $\sN_L,a_L\ge0$ and $G\neq\Delta$, there must exist $s>0$ such that $G(s)<1$. It then follows that there exists $\varepsilon<1$ and
  $\alpha\in(0,s)$ such that $G$ is continuous at $\alpha$ and $G(\alpha)<\varepsilon$, which implies
  \begin{equation}
  \lim_{L\to\infty}\PP\left(\sN_L/a_L \le \alpha\right) = G(\alpha) < \varepsilon.
  \end{equation}
  Therefore there exists $\alpha>0$ and $\epsilon<1$ such that for all sufficiently large $L$
  \begin{equation}
  \PP(\sN_L/a_L \le \alpha) < \epsilon.
  \end{equation}
  Now let $m_L:=\lfloor\alpha a_L\rfloor - \ind(x_L\nleftrightarrow\lfloor\alpha a_L\rfloor)$. Then $m_L\ge0$ for all sufficiently large
  $L$, and 
  \begin{equation}
    \label{penultimate lower bound on infinite lattice g}
    g_{\sN_L}(x_L) \ge p_{m_L}(x_L)\,\PP(\sN_L\ge m_L) \ge (1-\epsilon) p_{m_L}(x_L). 
  \end{equation}

  By construction, $x_L\leftrightarrow m_L$ and $m_L\to\infty$ as $L\to\infty$, and so it follows from~\cite[Theorem 1.2.1,
    (1.10)]{Lawler91} that there exists $C\in(0,\infty)$ such that for all sufficiently large $L$
  \begin{equation}
    p_{m_L}(x_L)\ge m_L^{-d/2}\left[2\left(\frac{d}{2\pi}\right)^{d/2}\exp\left(-\frac{d}{2}\frac{|x_L|^2}{m_L}\right)-\frac{C}{m_L}\right]
  \end{equation}
  But $|x_L|^2/m_L = (\xi^2/\alpha)[1+o(1)]$ and so 
  \begin{equation}
    \label{lower bound on p_{m_L}(x_L)}
    p_{m_L}(x_L) \gg_\xi \,a_L^{-d/2}\;.
  \end{equation}
  Combining~\eqref{lower bound on p_{m_L}(x_L)} with~\eqref{penultimate lower bound on infinite lattice g} then yields the stated result.
\end{proof}

\section{Proof of Proposition~\ref{theorem:two-point function plateau}}
\label{sec:proof plateau}
For integer $n\ge1$, let 
\begin{equation}
\bar{p}_n(x ):= 2\left(\dfrac{d}{2\pi n}\right)^{d/2} \exp\left(-\frac{d\, |x|^2}{2n}\right)\;, \ \forall \ x \in \ZZ^d.
\label{Eq: pnbar(x)}
\end{equation}
The local central limit theorem for simple random walk approximates $p_n$ via $\pnbar$. 

We will make use of the following lemma, whose proof is deferred to the end of this section. Let
\begin{equation}
\Gamma(s,z):=\int_{z}^\infty\, t^{s-1}\,\e^{-t}\, \dt\;,\qquad z\ge0\;, 
\end{equation}
denote the (upper) incomplete gamma function, and let
 \begin{equation}
 \erfc(z):=\frac{2}{\sqrt{\pi}}\,\int_{z}^\infty\, \e^{-t^2}\, \dt\;,\qquad z\in\RR\;,
 \end{equation}
 denote the complementary error function. 
\begin{lemma}
\label{lemma:bound theta function}
Let $d\ge 1$ be an integer, let $a>0$ and let $b\in\RR^d$. Then,
\begin{equation}
  \label{eq:bound theta function sharp constants}
  \frac{\pi^{d/2}}{a^{d/2}}\,\erfc^d(2\sqrt{a}) - 1 \le \sum_{z\in\ZZ^d\setminus0} \e^{-a|z+b|^2}
  \le \frac{\pi^{d/2}}{a^{d/2}}\, \left(1+3\sqrt{\frac{a}{\pi}}\right)^d
\end{equation}
and there exist $c_d,C_d>0$ such that
\begin{equation}
  \label{eq:bound theta function generic}
c_d\, a^{-d/2} \Gamma(d/2,da) \le \sum_{z \in \ZZ^d\setminus0} \e^{-a|z|^2} \le C_d\, a^{-d/2}\;.
\end{equation}
\end{lemma}

\begin{proof}[Proof of Proposition~\ref{theorem:two-point function plateau}]
  Rearranging~\eqref{eq:concrete torus decomposition} yields
  \begin{equation}
  g_{\rP,L,\sN_L}(x_L ) =
  g_{\sN_L}(x_L ) + A(\sN_L,x_L,L) + E(\sN_L,x_L,L) + F(\sN_L,x_L,L)
  \label{eq:writing g torus in terms of E and F}
\end{equation}
where 
\begin{align}
  A(\sN,x,l) &:= \frac{1}{2}\sum_{z\in\ZZ^d\setminus0} \,\sum_{n=1}^\infty \pnbar(x +l\,z)\PP(\sN \ge n)\\
  E(\sN,x,l) &:= \sum_{z\in\ZZ^d\setminus0}\,\sum_{n=1}^{\infty}\PP(\sN\ge n)
  \frac{\pnbar(x + l\,z)}{2} [\ind(n\leftrightarrow x + l\,z) - \ind(n\nleftrightarrow x + l\,z)] \\
  F(\sN,x,l) &:= \sum_{z\in\ZZ^d\setminus0}\,\sum_{n=1}^{\infty}\PP(\sN\ge n)
  [p_n(x +l\,z) - \pnbar(x + l\,z)]\ind(n\leftrightarrow x + l\,z)
\end{align}
Similarly, rearranging~\eqref{eq:reflecting/holding to finite decomposition} yields, for $\ast=R,H$ and with $\tL=L$ if $\ast=\rH$ and
$\tL=L-1$ if $\ast=\rR$,
\begin{equation}
  \begin{split}
  g_{\ast,\,L,\,\sN_L}(x_L ) &=
  g_{\sN_L}(x_L) + B(L)+\tE(L)+\tF(L) \\
  &\qquad + \sum_{y\in[x_L]_{\tL}}\left(A(\sN_L,y,2\tL)+E(\sN_L,y,2\tL)+F(\sN_L,y,2\tL)\right)
  \label{eq:writing g box in terms of E and F}
  \end{split}
  \end{equation}
  where
\begin{align}
 B(L) &:= \frac{1}{2}\sum_{y\in[x_L]_{\tL}\setminus x_L}\,\sum_{n=1}^\infty\,\PP(\sN_L\ge n)\,\pnbar(y)\\
  \tE(L) &:= \sum_{y\in[x_L]_{\tL}\setminus x_L}\,\sum_{n=1}^{\infty}\PP(\sN_L\ge n)
  \frac{\pnbar(y)}{2} [\ind(n\leftrightarrow y) - \ind(n\nleftrightarrow y)] \\
  \tF(L) &:= \sum_{y\in[x_L]_{\tL}\setminus x_L}\,\sum_{n=1}^{\infty}\PP(\sN_L\ge n)
  [p_n(y) - \pnbar(y)]\ind(n\leftrightarrow y).
\end{align}

It is convenient in what follows to define the sequence $l_L$ via
\begin{equation}
  l_L=
  \begin{cases}
    L, & \ast = \rP,\\
    2L, & \ast = \rH,\\
    2(L-1), & \ast = \rR.
  \end{cases}
\end{equation}
and to then let $y_L$ denote an arbitrary sequence in $\ZZ^d$ satisfying $y_L\in\BB_{l_L}^d$. With this notation, we introduce the
abbreviations 
\begin{equation}
  A(L):= A(\sN_L,y_L,l_L), \quad E(L):=E(\sN_L,y_L,l_L),\qquad F(L):=F(\sN_L,y_L,l_L)\;.
\end{equation}

The first task is to show that $E(L)$, $\tE(L)$, $F(L)$ and $\tF(L)$ are all $o(\EE(\sN_L)/L^d)$.
We begin by considering $E(L)$ and $\tE(L)$. Since $\ind(w\nleftrightarrow n)=\ind(w \leftrightarrow n\pm 1)$, for any $w
\in\ZZ^d$
\begin{multline}
  \sum_{n=1}^\infty \bar{p}_n(w)\PP(\sN_L \ge n) \ind(n\leftrightarrow w) - \sum_{n=1}^\infty \bar{p}_n(w)\PP(\sN_L \ge n)
  \ind(n\nleftrightarrow w) \\
    \le \bar{p}_1(w ) + \sum_{n=1}^\infty \left|\bar{p}_{n+1}(w) - \bar{p}_n(w)\right|\,\PP(\sN_L\ge n),
\end{multline}
via a change of variables $n\mapsto n-1$ in the first sum. Similarly, changing variables in the second sum yields
\begin{multline}
  \sum_{n=1}^\infty \bar{p}_n(w)\PP(\sN_L \ge n) \ind(n\leftrightarrow w) - \sum_{n=1}^\infty \bar{p}_n(w)\PP(\sN_L \ge n)
  \ind(n\nleftrightarrow w) \\ 
\ge -\left(\bar{p}_1(w) + \sum_{n=1}^\infty \left|\bar{p}_{n+1}(w) - \bar{p}_n(w)\right|\,\PP(\sN_L\ge n)\right).  
\end{multline}
It then follows that 
\begin{align}
  \label{eq:initial bound for E}
  |E(L)| &\le \frac{1}{2}\sum_{z\in\ZZ^d\setminus0} \Bigg(\bar{p}_1(y_L +l_L\,z)+
   \sum_{n=1}^\infty |\bar{p}_{n}(y_L +l_L\,z) - \bar{p}_{n+1}(y_L + l_L\,z)|\,\PP(\sN_L\ge n) \Bigg)\\
  %% |E(L)| &\le \frac{1}{2}\sum_{z\in\ZZ^d\setminus0} \Bigg(\bar{p}_1(y_L +l_L\,z)  \nonumber\\
  %% & \qquad\qquad +
  %%  \sum_{n=1}^\infty |\bar{p}_{n}(y_L +l_L\,z) - \bar{p}_{n+1}(y_L + l_L\,z)|\,\PP(\sN_L\ge n) \Bigg)\\
   |\tE(L)| &\le \frac{1}{2}\sum_{y\in[x_L]_{\tL}\setminus x_L} \left(\bar{p}_1(y) +
   \sum_{n=1}^\infty |\bar{p}_{n}(y) - \bar{p}_{n+1}(y)|\,\PP(\sN_L\ge n)\right)
  \label{eq:initial bound for Etilde}   
\end{align}

Now define $\delta:=(1+d/2)^{-1}$; the motivation for this will become clear following~\eqref{eq:Markov bound for tail terms}. We first
consider the small $n$ terms in~\eqref{eq:initial bound for E} and~\ref{eq:initial bound for Etilde}. Let $w,z\in\ZZ^d\setminus 0$, with
$|w|\ge |z|\,l_L/4$. If $1\le n\le \lceil |z l_L|^{2-\delta}\rceil$ then for all $1\le j \le n+1$ we have 
\begin{equation}
 \label{eq:pnbar bound for small n}
 \bar{p}_j(w) \ll \e^{-\kappa\, l_L^{\delta} |z|^{\delta}}
 \end{equation}
with $\kappa=\kappa(d)>0$. The triangle inequality then implies that
 \begin{equation}
   \bar{p}_1(w) + \sum_{n=1}^{\ceil{|z\, l_L|^{2-\delta}}}\,|\bar{p}_n(w)-\bar{p}_{n+1}(w)|\,\PP(\sN_L\ge n)
   \;\ll\;
   |z\, l_L|^{2-\delta}\, \e^{-\kappa\,l_L^\delta\,|z|^{\delta}} \\
   \label{eq:single term small n bound for E}
 \end{equation}
Since $|y|\ge l_L/4$ for all $y\in[x_L]_{\tL}\setminus x_L$ and $x_L\in\boxL$, and since $[x_L]_{\tL}$ has only $2^d$ terms, it immediately
follows from~\eqref{eq:single term small n bound for E} with $z=1$ that we can bound the small $n$ terms in~\eqref{eq:initial bound for
  Etilde} via
\begin{equation}
  \sum_{y\in[x_L]_{\tL}\setminus x_L}\left( \bar{p}_1(y) + \sum_{n=1}^{\ceil{l_L^{2-\delta}}}\,|\bar{p}_n(y)-\bar{p}_{n+1}(y)|\;\PP(\sN_L\ge
  n)\right) 
  \ll l_L^{2-\delta} \e^{-\kappa\,l_L^{\delta}}\;.
  \label{eq:Etilde small n bound}
\end{equation}

But since $l_L^{\delta} |z|^{\delta} -l_L^{\delta}-|z|^{\delta}+1=(l_L^{\delta}-1)(|z|^{\delta}-1)\ge0$ for any $z\in\ZZ^d\setminus 0$, we have
that
\begin{equation}
  \sum_{z \in \mathbb{Z}^d \setminus 0 } |z\,l_L|^{2-\delta}\;\e^{-\kappa\, l_L^{\delta}|z|^\delta}
  \ll \; l_L^{2-\delta}\,\e^{-\kappa\,l_L^{\delta}}
  \label{eq:exp sum bound}
\end{equation}
Therefore, combining the fact that Lemma~\ref{lemma:remove x in (x+zL)^2} implies $|y_L+l_L\, z|\ge |z|
l_L/2$, with~\eqref{eq:single term small n bound for E} and~\eqref{eq:exp sum bound}, it then follows that 
\begin{multline}
\label{eq:E1 early terms}
  \sum_{z \in \mathbb{Z}^d \setminus 0}
  \left(\bar{p}_1(y_L+l_L\,z)
  + \sum_{n=1}^{\ceil{|z l_L|^{2-\delta}}} \left\lvert \bar{p}_n(y_L + l_L\,z) - \bar{p}_{n+1}(y_L + l_L\,z)\right\rvert\PP(\sN_L\ge n) \right)
  \\  
  \ll \; l_L^{2-\delta}\,\e^{-\kappa\, l_L^\delta}.
\end{multline}
We therefore see that the sums of the small $n$ terms in $E(L)$ and $\tE(L)$ are exponentially small in $L$.

We now consider the large $n$ terms in~\eqref{eq:initial bound for E} and~\eqref{eq:initial bound for Etilde}. An elementary argument
shows that there exists $c_d>0$ such that for any $n\in\posint$ and $y\in\ZZ^d$ 
\begin{equation}
|\bar{p}_n(y) - \bar{p}_{n+1}(y)| \le c_d\, n^{-1-d/2}\;
\end{equation}
Using Markov's inequality, it then follows that for any $y,z\in\ZZ^d$
\begin{equation}
  \begin{split}
\sum_{n=\ceil{|z\, l_L|^{2-\delta}}+1}^{\infty} \left\lvert \bar{p}_n(y) - \bar{p}_{n+1}(y) \right\rvert \PP(\sN_L \ge n)
&\;\ll\; \sum_{n=\ceil{|z\, l_L|^{2-\delta}}+1}^{\infty} \frac{1}{n^{d/2+1}} \PP(\sN_L \ge n)\\
&\;\ll\; \EE(\sN_L) \, \sum_{n=\ceil{|z\,l_L|^{2-\delta}}+1}^{\infty} \frac{1}{n^{d/2+2}} \\
&\;\ll\; \EE(\sN_L) \, \int_{|z\, l_L|^{2-\delta}}^\infty \,t^{-d/2 - 2} \,\dt \\
&\;\ll\; \frac{1}{|z|^{d+1}}\,\frac{\EE(\sN_L)}{l_L^{d+1}}\;.
\label{eq:Markov bound for tail terms}
  \end{split}
  \end{equation}
Since $|z|^{-d-1}$ is summable over $\ZZ^d\setminus 0$, combining~\eqref{eq:E1 early terms}, \eqref{eq:Etilde small n bound}
and~\eqref{eq:Markov bound for tail terms} shows that
\begin{equation}
    \label{eq:final E bound}
    \tE(L),\quad E(\sN_L,x_L,L),\quad \sum_{y\in[x_L]_{\tL}} E(\sN_L,y,2\tL)\, = O\left(\frac{\EE(\sN_L)}{L^{d+1}}\right).
\end{equation}

The bounds for $F(L)$ and $\tF(L)$ are obtained similarly, with the aid of the local central limit theorem. If $n\le \ceil{|z
  l_L|^{2-\delta}}$ and $w,z\in\ZZ^d\setminus 0$ satisfy $|w|\ge |z|\,l_L/4$, then it follows from~\cite[Proposition 2.1.2]{LawlerLimic2010}
that 
\begin{equation}
p_n(w) \le \PP\left(\max_{0\le i \le n}|S_i| \ge |w|\right)
\ll \, \e^{-\varphi\, |z\, l_L|^{\delta}},
\label{eq:pn bound small n}
\end{equation}
for some $\varphi=\varphi(d)>0$. It then follows from~\eqref{eq:pn bound small n}, \eqref{eq:pnbar bound for small n} and the
triangle inequality that 
\begin{equation}
  \sum_{n=1}^{\ceil{|z\,l_L|^{2-\delta}}} \, |p_n(w) - \bar{p}_n(w)|\,\PP(\sN_L\ge n)\ind(n\leftrightarrow w)
  \;\ll\;
  |z\,l_L|^{2-\delta} \e^{-(\kappa\wedge\varphi)\,l_L^{\delta}\,|z|^{\delta}}.
  \label{eq:F preliminary small n bound}
\end{equation}
Just as~\eqref{eq:Etilde small n bound} and~\eqref{eq:E1 early terms} were obtained from~\eqref{eq:single term small n bound for E}
and~\eqref{eq:exp sum bound}, analogous arguments using~\eqref{eq:F preliminary small n bound} yield 
\begin{equation}
  \sum_{y\in[x_L]_{\tL}\setminus x_L}\sum_{n=1}^{\ceil{l_L^{2-\delta}}} \abs{p_n(y) - \bar{p}_n(y)} \PP(\sN_L\ge n)
  \ind(n\leftrightarrow y)
  \ll\; l_L^{2-\delta}\,\e^{-(\kappa\wedge\varphi)\, l_L^{\delta}}\;,
  \label{eq:Ftilde early terms}
\end{equation}
and
\begin{equation}
  \sum_{z\in\ZZ^d\setminus0}\!\sum_{n=1}^{\ceil{|z\, l_L|^{2-\delta}}} \abs{p_n(y_L  + l_L z) - \bar{p}_n(y_L  + l_L z)} \PP(\sN_L\ge n)
  \ind(n \leftrightarrow y_L+l_L z)
  \ll\; l_L^{2-\delta}\,\e^{-(\kappa\wedge\varphi)\, l_L^{\delta}}
\label{eq:F early terms}
\end{equation}

The local central limit theorem for simple random walk (see e.g.~\cite[Theorem 1.2.1]{Lawler2013}) implies that that for all
$y\in\ZZ^d$ 
\begin{equation}
  |p_n(y) - \bar{p}_n(y)|\,\PP(\sN_L\ge n)\ind(n\leftrightarrow y)
  \; \ll \; n^{-1-d/2}\;.
\end{equation} 
Arguing as in~\eqref{eq:Markov bound for tail terms}, and combining with~\eqref{eq:Ftilde early terms} and~\eqref{eq:F early terms} then
yields 
\begin{equation}
  \label{eq:final F bound}
    \tF(L),\quad F(\sN_L,x_L,L),\quad \sum_{y\in[x_L]_{\tL}} F(\sN_L,y,2\tL)\, = O\left(\frac{\EE(\sN_L)}{L^{d+1}}\right).
\end{equation}

Next we consider $B(L)$. Let $a_L$ be a positive sequence. For any $y\in\ZZ^d$ we have
\begin{equation}
  \begin{split}
    \sum_{n=\ceil{L^2\,a_L}}^\infty\;\PP(\sN_L\ge n)\; \frac{\pnbar(y)}{2}
    &\ll \; \frac{1}{a_L^{d/2}\,L^d}\;\sum_{n=\ceil{L^2\,a_L}}^\infty\;\PP(\sN_L\ge n)\\
    &\ll a_L^{-d/2}\,\frac{\EE(\sN_L)}{L^d}\;.
    \label{eq:B large n}    
  \end{split}
\end{equation}
  If now $|y|\ge \tL/2$, then since $\e^{-t}<t^{-\gamma}$ for all $t,\gamma>0$, taking $\gamma=d/2$ we have
  \begin{equation}
    \frac{\pnbar(y)}{2} \ll \;n^{-d/2}\;\e^{-d\,\tL^2/(8n)} \ll L^{-d}
  \end{equation}
  and therefore
  \begin{equation}
    \sum_{n=1}^{\ceil{L^2\,a_L}-1}\;\PP(\sN_L\ge n)\,\frac{\pnbar(y)}{2}
    \ll
    L^{2-d}\,a_L.
    \label{eq:B small n}
  \end{equation}
  Consequently, if $\EE(\sN_L)=\Omega(L^2)$ then choosing $a_L\to1$ in~\eqref{eq:B large n} and~\eqref{eq:B small n} implies
  \begin{equation}
    B(L) = O(L^{-d}\,\EE(\sN_L))\;,
    \label{eq:weak bound on B}
  \end{equation}
  while if $\EE(\sN_L)=\omega(L^2)$ then choosing $a_L\to\infty$ with $a_L L^2=o(\EE(\sN_L))$ implies
  \begin{equation}
    B(L) = o(L^{-d}\,\EE(\sN_L))\;.
    \label{eq:sharp bound on B}
  \end{equation}
  
  It now remains to study the asymptotic behaviour of $A(L)$. We first consider Part~\ref{eq:plateau theorem weak}, and therefore assume
  $\meanNL = \Omega(L^2)$. It follows from Lemma~\ref{lemma:remove x in (x+zL)^2} and~\eqref{eq:bound theta function generic} of
  Lemma~\ref{lemma:bound theta function} that 
  \begin{equation}
  A(\sN_L,y_L,l_L) \;\ll\; l_L^{-d} \sum_{n=1}^{\infty} \PP(\sN_L \ge n) \ll \frac{\meanNL}{L^{d}} \;.
  \label{eq:Upper bound for A}
\end{equation}

Since $\meanNL=\Omega(L^2)$, there exists $\gamma>0$ such that $\EE(\sN_L)\ge \gamma\,L^2$ for all sufficiently large $L$. It
follows again Lemma~\ref{lemma:remove x in (x+zL)^2} and~\eqref{eq:bound theta function generic} of Lemma~\ref{lemma:bound theta function}
that 
\begin{align}
  A(\sN_L,y_L,l_L)
&\gg  l_L^{-d}\,\sum_{n=1}^\infty \,\PP(\sN_L \ge n)\, \Gamma\left(\frac{d}{2},\frac{2\,d^3\, l_L^2}{n}\right)\\
& \gg L^{-d}\sum_{n=\floor{\gamma\,L^2/2}+1}^\infty \PP(\sN_L \ge n) \\
&= L^{-d} \left(\meanNL - \sum_{n=1}^{\floor{\gamma\,L^2/2}} \PP(\sN_L \ge n)\right) \\
& \gg \frac{\meanNL}{L^d}
\end{align}
It follows in particular that
\begin{equation}
   A(\sN_L,x_L,L) \asymp \frac{\meanNL}{L^d}\;,\qquad \sum_{y\in[x_L]_{\tL}}\,A(\sN_L,y,2\tL) \asymp \frac{\meanNL}{L^d}.
 \end{equation}
Together with~\eqref{eq:weak bound on B}, \eqref{eq:final E bound}, \eqref{eq:final F bound},
\eqref{eq:writing g torus in terms of E and F} and~\eqref{eq:writing g box in terms of E and F} this establishes Part~\ref{eq:plateau
  theorem weak}.

We now focus on Part~\ref{eq:plateau theorem sharp}, and therefore assume $\meanNL = \omega(L^2)$. Let $a_L>0$ be a sequence satisfying
$a_L\to\infty$ and $a_L = o(\meanNL/L^2)$. From Lemma~\ref{lemma:remove x in (x+zL)^2} and~\eqref{eq:bound theta function generic} of
Lemma~\ref{lemma:bound theta function} we have 
\begin{equation}
  \begin{split}
    \label{eq:A small n terms}
    \frac{1}{2}
    \sum_{n=1}^{\ceil{a_L L^2}-1} \PP(\sN_L\ge n)\,\sum_{z\in\ZZ^d\setminus0}\,\pnbar(y_L+l_L z)
    &\ll l_L^{-d}\,\sum_{n=1}^{\ceil{a_L L^2}-1} \PP(\sN_L\ge n)\\
    &\ll a_L\, L^{2-d} \\ &= o\left(\frac{\EE(\sN_L)}{L^d}\right)\;.
  \end{split}
\end{equation} 
But~\eqref{eq:bound theta function sharp constants} of Lemma~\ref{lemma:bound theta function} implies
\begin{equation}
  \begin{split}
    \frac{1}{2}\sum_{n=\ceil{a_L L^2}}^\infty \PP(\sN_L\ge n)\sum_{z\in\ZZ^d\setminus0} \pnbar(y_L+l_L z)
      &\;\le\; l_L^{-d}\sum_{n=\ceil{a_L L^2}}^\infty \PP(\sN_L\ge n) \left(1+3\sqrt{\frac{d\,l_L^2}{2\pi\,a_L\,L^2}}\right)^d\\
         &= l_L^{-d} [1+o(1)]\,\sum_{n=\ceil{a_L L^2}}^\infty \,\PP(\sN_L\ge n)\;.
  \end{split}
\end{equation}
and, since $\erfc$ is decreasing and continuous and $\erfc(0)=1$, 
\begin{align}
  \frac{1}{2}&\,\sum_{n=\ceil{a_L L^2}}^\infty \,\PP(\sN_L\ge n)\,\sum_{z\in\ZZ^d\setminus0} \pnbar(y_L+l_L z)
  \nonumber
  \\
  &\;\ge\; l_L^{-d}\,\sum_{n=\ceil{a_L L^2}}^\infty \,\PP(\sN_L\ge n)
  \left[\erfc^d\left(\sqrt{\frac{2\,d\,l_L^2}{a_L\,L^2}}\right)-\left(\frac{d\,l_L^2}{2\,\pi\,a_L\,L^2}\right)^{d/2}\right]\nonumber\\
  &= l_L^{-d} [1+o(1)]\,\sum_{n=\ceil{a_L L^2}}^\infty \,\PP(\sN_L\ge n)\;.
\end{align}
But, by assumption on $a_L$, 
\begin{equation}
    \sum_{n=1}^{\ceil{a_L\,L^2}-1}\,\PP(\sN_L\ge n) \le a_L\,L^2 = o(\EE(\sN_L))
\end{equation}
and so
\begin{equation}
  \label{eq:A big n terms}
  \frac{1}{2}\,\sum_{n=\ceil{a_L L^2}}^\infty \,\PP(\sN_L\ge n)\,\sum_{z\in\ZZ^d\setminus0} \pnbar(y_L+l_L z)
  =\frac{\EE(\sN_L)}{l_L^d}\,[1+o(1)]\;.
\end{equation}
Combining~\eqref{eq:A small n terms} and~\eqref{eq:A big n terms} we then have
\begin{equation}
  A(\sN_L,y_L,l_L) \sim \frac{\EE(\sN_L)}{l_L^d}\;.
  \label{eq:sharp A asymptotics}
\end{equation}
In particular, for periodic boundary conditions we have $l_L=L$ and so~\eqref{eq:sharp A asymptotics} implies
\begin{equation}
  A(\sN_L,x_L,L) \sim \frac{\EE(\sN_L)}{L^d}
  \label{eq:sharp A asymptotics periodic}
\end{equation}
while for reflecting or holding boundary conditions we have $l_L=2\tL$ and so~\eqref{eq:sharp A asymptotics} implies
\begin{equation}
  \label{eq:sharp A asymptotics box}  
  \sum_{y\in[x_L]_{\tL}}\, A(\sN_L,y,2\tL) \sim
  \sum_{y\in[x_L]_{\tL}}\, \frac{\EE(\sN_L)}{2^d\,L^d} 
  = \frac{\EE(\sN_L)}{L^d}\;.
\end{equation}
Together with~\eqref{eq:sharp bound on B}, \eqref{eq:final E bound}, \eqref{eq:final F bound},
\eqref{eq:writing g torus in terms of E and F} and~\eqref{eq:writing g box in terms of E and F} this establishes Part~\ref{eq:plateau
  theorem sharp}. 
\end{proof}

\begin{proof}[Proof of Lemma~\ref{lemma:bound theta function}]
Let $\epsilon\in(-1,1)$. For any $z\ge2$ we have $\e^{-a(z+\epsilon)^2} \le \e^{-a(t-1/2+\epsilon)^2} $ for all $t\in [z-1/2,z+1/2]$, and so
\begin{equation}
  \label{eq:posint sum upper bound}
  \sum_{z=1}^\infty \e^{-a (z+\epsilon)^2}
  \le e^{-a(1+\epsilon)^2} + \sum_{z=2}^\infty  \int_{z-1/2}^{z+1/2} \e^{-a(t-1/2+\epsilon)^2} \dt 
  \le 1+ \frac{1}{2}\sqrt{\frac{\pi}{a}}\;.
\end{equation}
A similar argument also produces the lower bound
\begin{equation}
  \label{eq:posint sum lower bound}
  \sum_{z=1}^\infty \e^{-a(z+\epsilon)^2} \ge \sum_{z=1}^\infty  \int_{z-1/2}^{z+1/2} \e^{-a(t+1/2+\epsilon)^2} \dt
  \ge \frac{1}{2}\,\sqrt{\frac{\pi}{a}}\,\erfc(2\sqrt{a})\;.
\end{equation}

Now let $\alpha\in\RR$, and define $\{\alpha\}:=\alpha-\floor{\alpha}\in[0,1)$. Since $z\mapsto z-\floor{\alpha}$ is a bijection of $\ZZ$,
we have
  \begin{equation}
    \sum_{z\in\ZZ}\, e^{-a(z+\alpha)^2} = \sum_{z\in\posint} e^{-a(z+\{\alpha\})^2} + \sum_{z\in\posint} e^{-a(z-\{\alpha\})^2} +
    e^{-a\{\alpha\}^2}
  \end{equation}
It follows from~\eqref{eq:posint sum upper bound} and~\eqref{eq:posint sum lower bound} that
  \begin{equation}
    \sqrt{\frac{\pi}{a}}\,\erfc(2\sqrt{a}) \le \sum_{z\in\ZZ} e^{-a(z+\alpha)^2} \le \sqrt{\frac{\pi}{a}} + 3
  \end{equation}
Eq.~\eqref{eq:bound theta function sharp constants} then follows by observing that
\begin{equation}
\sum_{z \in \ZZ^d\setminus0}\e^{-a|z+b|^2} = \prod_{i=1}^d \sum_{z_i\in\ZZ}\e^{-a (z_i+b_i)^2} - 1 \;.
\end{equation}

We now consider~\eqref{eq:bound theta function generic}. Let $n\in \posint$ and $\AA_n = [-n, n]^d \cap \mathbb{Z}^d$. Let $\partial
\AA_n = \AA_n \setminus \AA_{n-1}$ be the set of vertices on the surface of the box $\AA_n$. Since
$\AA_{n-1} \subset \AA_n$, it follows that 
\begin{equation}
\abs{\partial \AA_n} = |\AA_n| - |\AA_{n-1}| = (2n + 1)^d - (2n - 1)^d\;.
\end{equation}
Therefore there exist $c_d,C_d>0$ such that for all $n\in\posint$
\begin{equation}
  c_d\, n^{d-1} \le |\partial\,\AA_n| \le C_d\,n^{d-1}\;.
\end{equation}
Now observe that if $|z|_\infty$ denotes the sup norm on $\RR^d$, then $\partial\AA_n=\{z\in\ZZ^d:|z|_\infty =n\}$ and
$|z|_\infty \le |z| \le \sqrt{d}\,|z|_\infty$. Consequently, 
\begin{equation}
  \begin{split}
\sum_{z \in \ZZ^d\setminus0} \e^{-a\,|z|^2} & \le  \sum_{n=1}^\infty \e^{-a\,n^2} \abs{\partial \AA_n} \\
& \le C_d\,\sum_{n=1}^\infty \int_{n}^{n+1}\,\e^{-a\,t^2/4}\, t^{d-1}\,\dt \\
& \le \left(2^{d-1}\,\Gamma(d/2)\,C_d\right)\,a^{-d/2}
  \end{split}
  \end{equation}
A lower bound is obtained similarly:
\begin{equation}
  \begin{split}
\sum_{z \in \ZZ^d\setminus0} \e^{-a\,|z|^2} & \ge  \sum_{n=1}^\infty \e^{-a\, d\, n^2}\, \abs{\partial \AA_n}\\
& \ge c_d\,\sum_{n=1}^\infty \int_{n}^{n+1}\e^{-a\,d\, t^2}\, \left(\frac{t}{2}\right)^{d-1}\,\dt \\
& \ge \left(\frac{c_d}{2^d\,d^{d/2}}\right)\, a^{-d/2}\, \Gamma(d/2,d\,a)\;.
  \end{split}
  \end{equation}
\end{proof}

\section{Proof of Lemma~\ref{lemma:two-point function projection identities}}
\label{sec:proof of projection lemma}
Let $(X_t)_{t\in \NN}$ be a Markov chain on a countable set $S$ with transition matrix $P$. Let $\mathcal{N}$ be an $\mathbb{N}$-valued
random variable, independent of $(X_t)_{t\in\NN}$. If $X_0=x_0$ for some fixed $x_0\in S$, then we define the corresponding two-point
function to be 
\begin{equation}
\label{Def: two point function}
g(x_0,x): = \EE_{x_0}\left(\,\sum_{t=0}^{\sN} \,\ind(X_t=x)\right) = \sum_{t=0}^{\infty}P^t(x_0, x) \PP(\sN\ge t)\;;
\end{equation}
the expected number of visits to $x\in S$ by time $\sN$.

Now let $\sim$ denote an equivalence relation on $S$. For each $x\in S$, we let $[x]:= \{x'\in S: x' \sim x \}$
denote its equivalence class, and denote the set of all equivalence classes on $S$ by $S^{\#}:= \{[x]: x\in S \}$.
We say $P$ respects $\sim$ if $P(x,[y]) = P(x',[y])$ for all $x,y\in S$ and all $x^\prime \sim x$. 
If $P$ respects $\sim$, it is straightforward to show that the matrix $P^\#$ on $S^\#$ defined by
\begin{equation}
  P^\#([x],[y]):=P(x,[y]):=\sum_{y\in[y]}P(x,y)
\end{equation}
is stochastic, and that the process $([X_t])_{t\in\NN}$ is a Markov chain on $S^\#$ with transition matrix $P^\#$. See
e.g.~\cite[\S2.3.1]{LevinPeres2017}.

\begin{lemma}
\label{lemma:projected two-point function}
Let $S$ be a countable set, endowed with an equivalence relation respected by the stochastic matrix $P$.
Let $(X_t)_{t\in \mathbb{N}}$ be a Markov chain with transition matrix $P$, and let $\sN$ be an $\NN$-valued random variable,
independent of $(X_t)_{t\in\mathbb{N}}$. Let $g$ be the corresponding two-point function of $(X_t)_{t\in\NN}$, and $g^\#$ be the
corresponding two-point function of $([X_t])_{t\in\NN}$. Then for $[x],[y]\in S^\#$ and all $x'\in[x]$ we have 
$$
g^{\#}([x], [y]) = \sum_{y'\in [y]} g(x', y')\;.
$$
\end{lemma}

\begin{proof}
  Let $x,y\in S$. A simple induction on $t$ shows that $(P^{{\#}})^t([x],[y]) = P^{t}(x',[y])$ for all integers $t\ge 0$ and all $x'\in [x]$. 
  Therefore if $x^\prime \in [x]$, it follows that
\begin{equation}
g^{\#}([x], [y]) = \sum_{t=0}^\infty \,P^{t}(x',[y])\,\PP(\sN\ge\,t) = \sum_{y'\in [y]}\,\sum_{t=0}^\infty\, P^{t}(x',y')\,\PP(\sN\ge t) =
\sum_{y'\in [y]}\, g(x', y') \;.
\end{equation}
\end{proof}

\subsection{Proof of Lemma~\ref{lemma:two-point function projection identities}}
\label{subsec:proof of projection lemma}
The proof of Lemma~\ref{lemma:two-point function projection identities} relies on the following two results, whose proofs are deferred to
Section~\ref{subsec:projection lemmas}. Recall the definition of $[x]_L$ given in~\eqref{eq:torus to box equivalence relation}.
\begin{lemma}
  \label{lemma:torus SRW respects equivalence relation}
  Let $L\ge2$, and let $P_{\rP,2L}$ denote the transition matrix of simple random walk on $\doubleboxL$ with periodic boundary conditions. If
  $x, y \in\BB_{2L}^d$, then for all $x^\prime\in[x]_L$
  $$
  P_{\rP,2L}(x,[y]_L) = P_{\rP,2L}(x^\prime, [y]_L).
  $$
\label{lemma:srw_transition_matrix_symmetric}
\end{lemma}

\begin{lemma}
\label{lemma:boundaryconditions}
Let $L\ge 3$, and let $P_{\rP,L}$, $P_{\rH,L}$, and $P_{\text{R},L}$ denote the transition matrices of simple random walk on $\boxL$ with
periodic, holding and reflective boundary conditions, respectively. For any odd $L$ we have
\begin{equation}
\label{Eq:R_to_P}
P_{\text{R},L} (x, y)= P_{\text{P},2(L-1)}^{\#}([x]_{L-1},[y]_{L-1})
\end{equation}
\begin{equation}
\label{Eq:H_to_P}
P_{\text{H},L} (x, y)= P_{\text{P},2L}^{\#}([x]_{L},[y]_{L})
\end{equation}
for all $x,y \in \mathbb{B}_{L}$.
\end{lemma}

\begin{proof}[Proof of Lemma~\ref{lemma:two-point function projection identities}]
Let $P$ denote the transition matrix of simple random walk on $\ZZ^d$, and let $P_{\ast,L}$ denote the transition matrix of simple random
walk on $\boxL$ with $\ast$ boundary conditions, where $\ast$ can denote $\rP$, $\rR$ or $\rH$, for periodic, reflecting or holding,
respectively. Fix a random variable $\sN$, and let $g$ denote the two-point function defined in~\eqref{Def: two point function}
corresponding to $P$ and $\sN$. Likewise, let $g_{\ast,L}$ denote the two-point function corresponding to $P_{\ast,L}$ and $\sN$. 
The corresponding two-point functions defined in Section~\ref{sec:random walk models} are the specialisations of
$g$ and $g_{\ast,L}$ to the case $x_0=0$. We therefore freely omit the first argument of~\eqref{Def: two point function} when convenient, with the
understanding that in such instances it takes the value $0$.
  
We begin by proving Part~\ref{projection:infinite to torus}. Consider the equivalence relation on $\ZZ^d$ defined so that to each
$x\in\boxL$ there corresponds the equivalence class 
  \begin{equation}
    [x] := \{x+\,L\,z\,:\,z\in\ZZ^d\}.
  \end{equation}
It is straightforward to verify that $P$ respects this equivalence relation, and a simple calculation shows that
$P^\#([x],[y])=P_{\rP,L}(x,y)$ for all $x,y\in\boxL$. We therefore have for any $x\in\torus$ that
\begin{equation}
  \begin{split}
  g_{\rP,L}(x) &:= \sum_{t=0}^{\infty}\,P_{\rP,L}^t(0,x)\,\PP(\sN\ge t)\\
  &\phantom{:}= \sum_{t=0}^\infty\, (P^{\#})^t([0],[x])\PP(\sN\ge t)\\
  &\phantom{:}= g^{\#}([0],[x]) \\
  &\phantom{:}= \sum_{x'\in[x]}\,g(0,x') \\
  &\phantom{:}= \sum_{z\in\ZZ^d}\,g(x+zL),
  \end{split}
\end{equation}
where the penultimate step follows from Lemma~\ref{lemma:projected two-point function}. This establishes Part~\ref{projection:infinite to torus}.

Similarly, since $\boxL\subset \TT_{2(L-1)}^d$ we have for all $x\in\boxL$ that
\begin{equation}
  \begin{split}
    g_{\rR,L}(x)&:=\sum_{t=0}^\infty\,P_{\rR,L}^t(0,x)\,\PP(\sN\ge t)\\
    &\phantom{:}=\sum_{t=0}^\infty\,(P_{\rR,2(L-1)}^\#)^t([0]_{L-1},[x]_{L-1})\,\PP(\sN\ge t)\\
    &\phantom{:}=g^{\#}_{\rP,2(L-1)}([0]_{L-1},[x]_{L-1})\\
    &\phantom{:}=\sum_{x'\in[x]_{L-1}}\,g_{\rP,2(L-1)}(x')
  \end{split}
\end{equation}
where the second step follows from Lemmas~\ref{lemma:srw_transition_matrix_symmetric} and~\ref{lemma:boundaryconditions}, and the last step
follows from Lemma~\ref{lemma:projected two-point function}. This establishes Part~\ref{projection:torus to
  reflecting}. Part~\ref{projection:torus to holding} is proved similarly.
 \end{proof}

\subsection{Proof of Lemmas~\ref{lemma:srw_transition_matrix_symmetric} and~\ref{lemma:boundaryconditions}}
\label{subsec:projection lemmas}
Recall the definition of $[x]_L$ given in~\eqref{eq:torus to box equivalence relation}. If $x,y\in\doubletorus$ and $i\in[d]$, we will write
$y_i\sim x_i$ iff $y_i\in\{x_i,-L-x_i\}$, where we emphasise that addition and multiplication are modulo $2L$. If $y_i\sim x_i$ for all
$i\in[d]$, then we will also write $y\sim x$, so that $y\sim x$ iff $y\in[x]_L$.  To avoid confusion, in this section we will denote
adjacency between two vertices $x,y\in\torus$ by $x\leftrightarrow y$, where the value of $L$ will be clear from the context.
\begin{proof}[Proof of Lemma~\ref{lemma:srw_transition_matrix_symmetric}]
Simple random walk on $\doubleboxL$ with periodic boundary conditions corresponds to simple random walk on $\doubletorus$.
Let $x,y\in\doubletorus$. Then
\begin{equation}
  \label{eq:transition matrix on even torus}
P_{\rP,2L}(x,[y]_L) = \sum_{y' \in [y]_L} P_{\rP,2L}(x,y') = \frac{1}{2d}|N(x) \cap [y]_L|
\end{equation}
where $N(x):=\{y \in \doubletorus: x \leftrightarrow y\}$ is the set of neighbours of $x$ in $\doubletorus$. 

Let $x' \sim x$. Suppose $N(x')\cap[y]_L$ is nonempty, and let $z' \in N(x') \cap [y]_L$.
Then $z' = x' + \delta e^k$ for some $\delta \in \{-1,1\}$ and $1 \le k \le d$, where $e^k$ denotes the standard unit vector
along the $k$th coordinate axis. Consider
\begin{equation}
z = \begin{cases} x + \delta e^k, & x_k' = x_k \\
  x - \delta e^k, & x_k' \neq x_k \end{cases}
\label{eq:z definition in projection proof}
\end{equation}
Clearly, $z \in N(x)$. Moreover, for all $i \neq k$ we have $z_i = x_i \sim x_i' = z_i' \sim
y_i$ since $z' \in [y]_L$, so that $z_i \sim y_i$. Furthermore, if $x_k = x_k'$ then $z_k = x_k + \delta =
x_k' + \delta = z_k' \sim y_k$, while if $x_k \neq x_k'$ then $z_k = x_k - \delta = (-L-x_k')-\delta = -L -
(x_k' + \delta) \sim x_k' + \delta = z_k' \sim y_k$, so that in either case $z_k \sim y_k$. It then follows that
$z \sim y$ and so $z \in N(x) \cap [y]_L$. In particular, we see that $N(x')\cap[y]_L$ is nonempty iff $N(x)\cap[y]_L$ is nonempty.

Suppose now that $N(x)\cap[y]_L$ is nonempty, and define $f:N(x)\cap[y]_L\to N(x')\cap[y]_L$ via
\begin{equation}
  f(x+\delta e^k) = \begin{cases}
    x'+\delta e^k, & x_k'=x_k\\
    x'-\delta e^k, & x_k'\neq x_k
    \end{cases}
\end{equation}
The above argument implies that for any $z'=x'+\delta e^k\in N(x')\cap[y]_L$, if we define $z\in N(x)\cap[y]_L$ as in~\eqref{eq:z definition
  in projection 
  proof} then 
\begin{equation}
  \begin{split}
f(z) &= \begin{cases} f(x+\delta e^k), & x_k' = x_k \\
f(x-\delta e^k), & x_k' \neq x_k \end{cases} \\
&= x' + \delta e^k \\
&= z'
\end{split}
\end{equation}
and so $f$ is surjective. 

Now suppose $z,z' \in N(x) \cap [y]_L$ satisfy $f(z)=f(z')$. Without loss of generality, suppose $z = x +\delta e^k$, so that 
\begin{equation}
\big( f(z) \big)_k = \begin{cases} x_k' +\delta, & x_k' = x_k \\
x_k' - \delta, & x_k' \neq x_k \end{cases}
\end{equation}

If $z' = x +\delta' e^{k'}$ with $k' \neq k$, then $\big( f(z') \big)_k = x_k' \neq x_k' \pm \delta$. So $f(z) = f(z')$ implies $k' = k$,
and it follows that
\begin{equation}
\big(f(z')\big)_k = \begin{cases} x_k' + \delta', & x_k' = x_k \\
x_k' - \delta', & x_k' \neq x_k \end{cases}
\end{equation}
and $\big(f(z) \big)_k = \big(f(z') \big)_k$ implies $\delta = \delta'$. Therefore, $z' =
x + \delta e^k = z$, which implies $f$ is injective. We conclude that since $f$ bijective, $|N(x) \cap [y]_L| = |N(x') \cap [y]_L|$ for all
$x,y\in\doubletorus$ and $x'\in[x]_L$. The stated result then follows from~\eqref{eq:transition matrix on even torus}. 
\end{proof}

\begin{proof}[Proof of Lemma~\ref{lemma:boundaryconditions}]
Let $l\in\posint$ and set $L=2l+1$. We first consider Eq.~\eqref{Eq:R_to_P}. By construction, $\BB_{2l+1}^d\subset\TT_{4l}^d$. Let
$S_{4l}^{\#}$ denote the set of equivalence classes on $\TT_{4l}^d$ defined by~\eqref{eq:torus to box equivalence relation}. Since the map
$\BB_{2l+1}^d \to S_{4l}^{\#}$ defined by $x\mapsto [x]_{2l}$ is a bijection, and since both $P_{\rR,2l+1}$ and $P_{\rP,4l}^\#$ are
stochastic, in order to show that $P_{\rR,2l+1}(x, y) = P_{\rP,4l}^\#([x]_{2l},[y]_{2l})$ for all $x, y \in \BB_{2l+1}^d$, it suffices to
consider only pairs $x,y\in \BB_{2l+1}^d$ with $P_{\rR,2l+1}(x,y) >0$. By definition, $P_{\rR,2l+1}(x,y) >0$ only if $y =x + \delta e^k$
for some $\delta \in \{-1,1\}$ and $k \in [d]$. 

Let $x \in \TT_{4l}^d$ with $-l \le x_i \le l$ for all $i \in[d]$. Suppose $y = x +\delta e^k$ for $\delta \in \{-1,1\}$ and
$k\in[d]$. Clearly, $y \in N(x) \cap [y]_{2l}$, where $N(x)$ is the set of neighbours of $x$ in $\TT_{4l}^d$.
Suppose $y' \in [y]_{2l}$. Then either $y_k' = y_k = x_k + \delta$ or $y_k' = -2l -y_k
=-2l-(x_k + \delta)$. Clearly $x_k +\delta \neq x_k$, and $-2l-(x_k+\delta)=x_k$ iff $-2l-\delta =2x_k$, but the latter cannot hold,
since the left-hand side is odd, the right-hand side is even, and addition is modulo $4l$. We therefore see that $y_k' \neq x_k$. In order
for $y'$ to belong to $N(x)$, it is therefore necessary that $y_i' = x_i$ for all $i \neq k$. Defining $\tilde{y}$ via $\tilde{y}_i = x_i$
for $i\neq k$ and $\tilde{y}_k=-2l-(x_k+\delta)$ we conclude that $\{y\} \subset N(x) \cap [y]_{2l} \subset \{y,\tilde{y} \}$. Since
$\tilde{y} \in [y]_{2l}$ by construction, we have
\begin{equation}
N(x) \cap [y]_{2l} = \begin{cases} \{y, \tilde{y} \}, & \tilde{y} \in N(x) \\
\{y \}, & \tilde{y} \notin N(x)\end{cases}
\end{equation}
But $\tilde{y} \in N(x)$ if and only if $\tilde{y}_k = x_k + \epsilon$ for $\epsilon \in \{-1,1 \}$. And
$\tilde{y} \neq y$ if and only if $y_k' \neq y_k$. So $|N(x) \cap [y]| = 2$ if and only if
$\tilde{y}_k = x_k - \delta$, which holds iff $-2l -(x_k + \delta) = x_k - \delta$ modulo $4l$, which in turn holds iff $x_k \in \{-l,
l\}$. It follows that if $y = x + \delta e^k$ then
\begin{equation}
  \begin{split}
P^{\#}_{\rP,4l}([x]_{2l},[y]_{2l}) =  P_{\rP,4l}(x,[y]) = \frac{1}{2d} |N(x) \cap
[y ]| &= \begin{cases} 1/(2d), & x_k \neq \pm l \\ 
1/d, & x_k = \pm l \end{cases} \\
&= P_{\rR,2l+1}(x,y)
\end{split}
\end{equation}

We next consider Eq.~\ref{Eq:H_to_P}. Note that $\BB_{2l+1}^d \subset \TT_{2(2l+1)}^d$, and let $S_{2(2l+1)}^{\#}$ denote the set of
equivalence classes on $\TT_{2(2l+1)}^d$ corresponding to~\eqref{eq:torus to box equivalence relation}. Since the map $\BB_{2l+1}^d \to
S_{2(2l+1)}^{\#}$ defined by $x \mapsto [x]_{2l+1}$ is a bijection, arguing as above it suffices to show
$P_{\rP,2(2l+1)}^{\#}([x]_{2l+1},[y]_{2l+1})=P_{\rH,2l+1}(x,y)$ for all $x,y\in\BB_{2l+1}^d$ with $P_{\rH,2l+1}(x,y)>0$. By definition,
$P_{\rH,2l+1}(x,y)>0$ only if $y = x + \delta e^k$ for some $\delta \in \{-1,1\}$ and $k \in[d]$ or if $y = x$. 

Let $x \in \BB_{2l+1}^d$. It is straightforward to show that $x + \delta e^k \sim x$ if and only if $x_k = \delta l$ which implies $|N(x)
\cap [x]_{2l+1}| = \sum\limits_{i=1}^d \ind \big(|x_i| = l \big)$, where $N(x)$ is the set of neighbours of $x$ in $\TT_{2(2l+1)}^d$, and therefore
\begin{equation}
P_{\rP,2(2l+1)}^{\#}([x]_{2l+1},[x]_{2l+1}) = P_{\rP,2(2l+1)}(x,[x]_{2l+1})=\frac{1}{2d} \sum_{i=1}^d \ind\Big(|x_i|= l \Big) = P_{\rH,2l+1}(x,x).
\end{equation}
Suppose instead that $y = x + \delta e^k$ with $x_k \neq \delta l$. Then $y \not\sim x$ and so if $y' \in [y]_{2l+1}$ then $y' \neq x$. If
$y' \in [y]_{2l+1}$ then either $y_k' = y_k = x_k + \delta$ or $y_k' = -2l-1-y_k$. But since $x_k \in \{-(l-1),...,(l-1)\}$, we have
$x_k+\epsilon\in\{-l,...,l\}$ for $\epsilon\in\{-1,1\}$ and
\begin{equation}
  -2l-1-y_k \in \{-2l-1,\ldots,-l-1\}\cup\{l+1,\ldots,2l\}\;.
\end{equation}
It follows that $-2l-1-y_k\neq x_k,x_k+\epsilon$. But in order for $y'\in N(x)$, it is necessary that $y_k'\in\{x_k,x_k\pm 1\}$. So
we conclude that if $y'\in N(x)$ then $y_k'\neq -2l-1-y_k$, which then implies that $y_k'=y_k=x_k+\delta$. It also then follows that
$y_i' = x_i =y_i$ for all $i \neq k$, so that $y' = y$. We have therefore established that $N(x) \cap [y]_{2l+1} = \{y\}$ when $y = x +
\delta e^k$ with $x_k \neq \delta l$, and it follows that
\begin{equation}
  P_{\rP,2(2l+1)}^{\#}([x]_{2l+1},[y]_{2l+1}) = P_{\rP,2(2l+1)}(x,[y]_{2l+1}) = 1/(2d)  = P_{\rH,2l+1}(x,y).
\end{equation}
\end{proof}

\ack{
 The authors thank Gordon Slade for sharing an earlier version of~\cite{Slade2023} during the final stages of completion of an earlier
 version of this work. This research was supported by the Australian Research Council Centre of Excellence for Mathematical and Statistical
 Frontiers (Project no. CE140100049), and the Australian Research Council's Discovery Projects funding scheme (Project
 No.s DP180100613 and DP230102209). It was undertaken with the assistance of resources and services from the National Computational
 Infrastructure (NCI), which is supported by the Australian Government. Y.D. was supported by the National Natural Science Foundation of
 China (under Grant No. 12275263), the Innovation Program for Quantum Science and Technology (under grant No. 2021ZD0301900), and the
 Natural Science Foundation of Fujian province of China (under Grant No. 2023J02032).
}
 
\section*{References}
\bibliographystyle{unsrt}

\end{document}